%% file: snam11.tex
\RequirePackage{fix-cm}

\documentclass[twocolumn]{svjour3}          

\smartqed  

\usepackage{epsfig}
\usepackage{amsmath,amssymb}
\usepackage{algorithm}
\usepackage[noend]{algorithmic}
\usepackage{multirow}
\usepackage{url}
\usepackage{natbib}
\usepackage{color}

\definecolor{Fra_color}{rgb}{0.2,0.6,0.2}

\definecolor{Amit_color}{rgb}{0.2,0.2,0.8}

\definecolor{Laks_color}{rgb}{0.8,0.0,0.5}

\def\0{{\mathbf 0}}
\def\1{{\mathbf 1}}
\def\C{{\mathcal C}}

\newcommand{\f}{\mathit{f}}
\newcommand{\g}{\mathit{g}}
\renewcommand{\S}{\mathit{S}}

\newcommand{\OPT}{\mathit{OPT}}

\newcommand{\MINSEEDSET}{\textsc{Greedy-Mintss}}

\newcommand{\RAKC}{\mathit{RAKC}}
\newcommand{\PSC}{\mathit{PSC}}
\newcommand{\X}{\mathcal{X}}
\newcommand{\I}{\mathcal{I}}
\newcommand{\U}{\mathcal{U}}
\newcommand{\J}{\mathcal{J}}
\newcommand{\A}{\mathcal{A}}
\newcommand{\ra}{\mbox{$\rightarrow$}}

\DeclareMathOperator*{\argmax}{arg\,max}
\newcommand{\reals}{\mathbb{R}}

\newcommand{\NPhard}{$\mathbf{NP}$-hard}

\newcommand{\SPhard}{$\mathbf{\#P}$-hard}

\newcommand{\meme}{\textsf{Meme}}
\newcommand{\neth}{\textsf{NetHEPT}}

\newcommand{\spara}[1]{\smallskip\noindent{\bf #1}}

\newcommand{\eat}[1]{}
\newcommand{\nop}[1]{}
\renewcommand{\cite}{\citet}

\begin{document}

\title{Approximation Analysis of Influence Spread in Social Networks}

\author{Amit Goyal \and Francesco Bonchi \and \\ Laks V. S. Lakshmanan
\and \\ Suresh Venkatasubramanian}


\institute{Amit Goyal and Laks V. S. Lakshmanan \at University of British Columbia,
Vancouver, BC, Canada. \\ \email{\{goyal, laks\}@cs.ubc.ca}  
\and 
Francesco Bonchi \at Yahoo! Research, Barcelona, Spain.
\\ \email{bonchi@yahoo-inc.com} 
\and 
Suresh Venkatasubramanian \at
University of Utah, Salt Lake City, UT, USA.
\\ \email{suresh@cs.utah.edu}
 }

\maketitle

\sloppy

\input{abstract}
\section{Introduction}
\label{sec:intro}
\input{new-intro}

\section{Preliminaries}
\label{sec:bg}
\input{background}

\section{Related Work}
\label{sec:related}
\input{related}

\section{Minimum Target Set Selection}
\label{sec:minseed}

\input{minseed}

\section{MINTIME}
\label{sec:mintime}
\input{mintime}

\section{Empirical Assessment}
\label{sec:exp}
\input{exp}

\section{Conclusions}
\label{sec:dis}
\input{discussion}


\small
\bibliographystyle{spbasic}
\bibliography{propagation}

\appendix

\normalsize
\section{Proof of Lemma~\ref{lem:set-tri}}
\label{sec:proof:lem:set-tri}

  Suppose there exists an algorithm $\A$ that selects $\beta k$ sets
  which covers $\gamma \eta$ elements. Apply $\A$ to an arbitrary
  instance $\langle \mathcal{U}, \mathcal{S}, \eta \rangle$ of $\PSC$.
  The output is a collection of sets $\C_1$ such that $|\C_1| \le
  \beta k$ and $|\bigcup_{\S \in \C_1} \S| \ge \gamma \eta$. Next,
  discard the sets that have been selected and the elements they
  cover, and apply again the algorithm $\A$ on the remaining universe.
  Repeat this process until $1$ or fewer elements are left
  uncovered.\footnote{Instead of $1$, we could be left with a constant
    number of elements. Asymptotically, it does not make a
    difference.}

  Let $\eta_i$ denote the number of elements uncovered after iteration
  $i$.  In iteration $i$, the algorithm picks $\beta k$ sets and
  covers at least $\gamma \eta_{i-1}$ elements. Hence, $\eta_i \le
  \eta_{i-1} \cdot (1 - \gamma)$.  Expanding, $\eta_i \le \eta \cdot
  (1 - \gamma)^i$. Suppose after $l$ iterations, $\eta_l = 1$. The
  total number of sets picked is $l\beta k$.  $\eta \cdot (1 -
  \gamma)^l = 1$ implies $l = \frac{\ln \eta}{\ln
    \frac{1}{1-\gamma}}$.

  We now prove the first claim. Let $\gamma > 1 - 1/e^\beta$, then
  $\ln \left( \frac{1}{1-\gamma} \right) > \beta$.  This yields a
  PTIME algorithm for PSC which outputs a solution of size $ l \beta k
  = \beta k \cdot \ln \eta / \ln \frac{1}{1-\gamma} \le c \cdot k \ln
  \eta$ (for some $c < 1$) This yields an $c \cdot \ln
  \eta$-approximation for PSC for some $c < 1$, which is not possible
  unless $NP \subseteq DTIME(n^{O(\log \log n)})$ \citep{feige98}.

  To prove the second claim, assume $\beta \le (1 - \delta) \ln \left(
    \frac{1}{1 -\gamma} \right)$.  This gives a PTIME algorithm for
  $\PSC$ which outputs a solution of size $l \beta k = \beta k \cdot
  \ln \eta / \ln \frac{1}{1-\gamma} \le (1 - \delta) k \cdot \ln \eta$
  which is not possible unless $NP \subseteq DTIME(n^{O(\log \log
    n)})$. 
  \qed
\section{Example Illustrating Performance of Wolsey's solution}
\label{sec:wolsey}

\cite{wolsey82} studied the RSSC problem and showed, among many things,
that the greedy algorithm provides a solution that is within a
factor of $1 + \ln (\eta/(\eta-f(S_{t-1}))$ of the optimal solution.
Unfortunately, this does not yield an approximation algorithm with any guaranteed bounds.
The following
example shows the greedy solution with threshold $\eta$
can be arbitrarily worse than the optimum.

\begin{figure}[htp]
\centering 
\includegraphics[width= 0.45\textwidth]{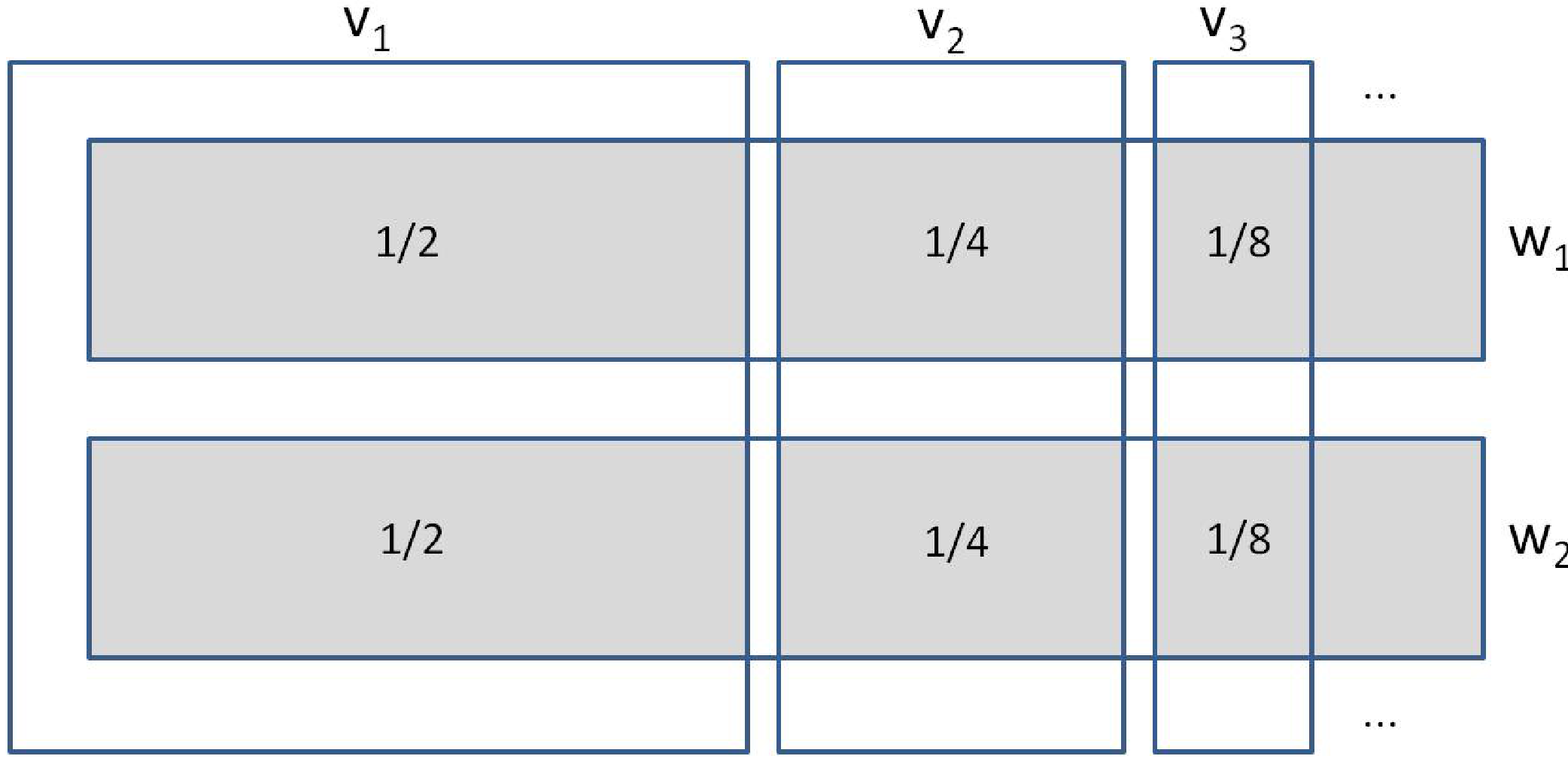}
\caption{Example. Rectangles represent the elements in the universe. The
shaded area within a rectangle represents the coverage function $\f$ for
the element. e.g., $f(v_1) = 1/2 + 1/2 = 1$.}
\label{fig:example}
\end{figure}

\noindent
\textbf{Example}
\label{ex:ce}(Illustrated also in Figure \ref{fig:example}).
Consider a ground set $\X = \{w_1, w_2, v_1, v_2, ..., v_l\}$ with
elements having unit costs. Figure~\ref{fig:example} geometrically
depicts the definition of a function $f: 2^{\X} \ra \reals$, where
for any set $S \subset \X$, $f(S)$ is defined to be the area
(shown shaded) covered by the elements of $S$. Specifically,
$f(w_1) = f(w_2) = 1 - 1/2^{l+1}$ and $f(v_i) = 1/2^{i-1}$, $1\le
i\le l$. Notice, $f(\{v_1, ..., v_l\}) = \Sigma_{i=1}^l 1/2^{i-1}
= 2 - 1/2^{l-1} < 2 - 1/2^l = f(\{w_1, w_2\})$. The greedy
algorithm will first pick $v_1$. Suppose it picks $S = \{v_1, ...,
v_i\}$ in $i$ rounds. Then $f(S\cup\{v_{i+1}\}) - f(S) = 1/2^i > 1
- 1/2^{l+1} - 1 + 1/2^i = 1 - 1/2^{l+1} - 1/2(2 - 1/2^{i-1}) =
f(S\cup\{w_1\}) - f(S)$. Thus, greedy will never pick $w_1$ or
$w_2$ before it picks $v_1, ..., v_l$. Suppose $\eta = 2 - 1/2^l$.
Clearly, the greedy solution is $\X$ whereas the optimal solution
is $\{w_1, w_2\}$. Here $l$ can be arbitrarily large.

\end{document}

%% file: abstract.tex
\begin{abstract}
In recent years, study of influence propagation in social networks
has gained tremendous attention.
In this context, we can
identify three orthogonal dimensions -- the number of \emph{seed}
nodes activated at the beginning (known as \emph{budget}), the
expected number of activated nodes at the end of the propagation
(known as \emph{expected spread} or \emph{coverage}), and the
\emph{time} taken for the propagation. We can constrain one or two
of these and try to optimize the third. In their seminal paper,
Kempe, Kleinberg and Tardos constrained the budget, left time
unconstrained, and maximized the
 cove\-rage: this problem is known as \emph{Influence Maximization} (or MAXINF for short).

In this paper, we study alternative optimization problems which are naturally
motivated by resource and time
constraints on viral marketing campaigns. In the first problem, termed \emph{Minimum
Target Set Selection} (or MINTSS for short), a coverage threshold $\eta$ is given and
the task is to find the \emph{minimum size seed set} such that by
activating it, at least $\eta$ nodes are eventually activated in
the expected sense. This naturally captures the problem of
deploying a viral campaign on a budget.
In the second problem, termed MINTIME, the goal is to minimize the time in which a predefined coverage is achieved.
More precisely, in  MINTIME, a coverage threshold $\eta$ and a budget threshold $k$
are given, and the task is to find a seed set of size at most $k$ such that by
activating it, at least $\eta$ nodes are activated in the expected
sense, \emph{in the minimum possible time}. This problem addresses
the issue of \emph{timing} when deploying viral campaigns. Both
these problems are \NPhard, which motivates our interest in their
approximation.

For MINTSS, we develop a simple greedy algorithm and show that it
provides a bicriteria approximation. We also establish a generic
hardness result suggesting that improving this bicriteria
approximation is likely to be hard. For MINTIME, we show that even
bicriteria and tricriteria approximations are hard under several
conditions. We show, however, that if we allow the budget for
number of seeds $k$ to be boosted by a logarithmic factor and
allow the coverage to fall short, then the problem can be solved
\emph{exactly} in PTIME, i.e., we can achieve the required
coverage within the time achieved by the optimal solution to
MINTIME with budget $k$ and coverage threshold~$\eta$.

Finally, we establish the value of the approximation algorithms, by
conducting an experimental evaluation, comparing their quality against
that achieved by various heuristics.

\keywords{Social Networks \and Social Influence \and Influence Propagation \and Viral Marketing \and Approximation
Analysis \and MINTSS \and MINTIME}

\end{abstract}

%% file: new-intro.tex
The study of how influence and information propagate in social
networks has recently received a great deal of attention
\citep{domingos01, domingos02, kempe03, kempe05, KimuraS06, leaders, ChenWY09, ChenWW10,
ChenWW10b, amit2010, weng2010, bakshy2011}.
One of the central problems in this domain is the problem of influence maximization
\citep{kempe03}.
Consider
a social network in which we have accurate estimates of influence
among users. Suppose we want to launch a new product in the market
by targeting a set of influential users (e.g., by offering them
the product at a discounted price), with the goal of starting a
word-of-mouth viral propagation, exploiting the power of social
connectivity. The idea is that by observing its neighbors
adopting the product, or more generally, performing an action, a
user may be influenced to perform the same action, with some
probability. Influence thus propagates in steps according to one
of the propagation models studied in the literature, e.g., the
\emph{independent cascade} (IC) or the \emph{linear threshold}
(LT) models ~\citep{kempe03}. The propagation stops when no new user
gets activated.

In this context, we can identify three main dimensions -- the
number of \emph{seed} nodes (or users) activated at the beginning
(known as the \emph{budget}), the expected number of nodes that
eventually get activated (known as \emph{coverage}  or
\emph{expected spread})\footnote{We use the terms coverage and expected spread
interchangeably throughout the article.}, and the number of \emph{time} steps
required for the propagation. In their seminal paper
\citet*{kempe03} 
 introduced the problem
of \textit{Influence Maximization} (MAXINF) which asks for a seed
set with a budget threshold $k$ that maximizes the expected spread
(time being left unconstrained). They showed that under the
standard propagation models IC and LT, MAXINF is \NPhard, but that
a simple greedy algorithm that exploits properties of the
propagation function yields a $(1-1/e -\phi)$-approximation, for
any $\phi>0$ (as discussed in detail in Section \ref{sec:bg}).

In this paper, we explore the other dimensions of influence propagation.
The problem of Minimum Target Set Selection (MINTSS) is motivated
by the observation that in a viral marketing campaign, we may be
interested in the smallest budget that will achieve a desired
outcome. The problem can therefore be defined as follows. We are
given a threshold $\eta$ for the expected spread and the problem
is to find a seed set of \emph{minimum size} such that activating
the set yields an expected spread of at least $\eta$.


In both MINTSS and MAXINF, the time for propagation is not
considered. Indeed, with the exception of a few papers
\citep[see e.g.,][]{LeskovecKDD07}, the temporal dimension of the social
propagation phenomenon has been largely overlooked. This is
surprising as the timeliness of a \emph{viral marketing} campaign
is a key ingredient for its success. Beyond viral marketing, many
other applications in time-critical domains can exploit social
networks as a means of communication to spread information
quickly.  This motivates the problem of Minimum Propagation Time
(MINTIME), defined as follows: given a
budget $k$ and a coverage threshold $\eta$, find a seed
set that satisfies the given budget and achieves the desired
coverage in \emph{as little time as possible}. Thus, MINTIME
tries to optimize the propagation time required to achieve a
desired coverage under a given budget.

\subsection{Our Contributions}

 We now summarize the main results in this paper.

 \begin{itemize}
\item  Firstly, we show (Section~\ref{sec:minseed}, Theorem~\ref{thm:sms_cost-ms})
that for all instances of MINTSS where the coverage function is
submodular, a simple greedy algorithm yields a bicriteria
approximation: given a coverage threshold $\eta$ and a shortfall
parameter $\epsilon > 0$, the greedy algorithm will produce a
solution $S$: $\sigma(S) \geq \eta - \epsilon$ and $|S| \leq (1 +
\ln(\eta/\epsilon)) OPT$, where $OPT$ is the optimal size of a
seed set whose coverage is at least $\eta$. That is, the greedy
solution exceeds the optimal solution in terms of size (budget) by
a logarithmic factor while achieving a coverage that falls short
of the required coverage by the shortfall parameter. We prove a
generic hardness result (Section ~\ref{sec:minseed},
Theorem~\ref{thm:inapprox3}) suggesting that improving this
approximation factor is likely to be hard. 

\smallskip

\item For MINTIME under IC and LT
model (or any model with monotone submodular coverage functions),
we show that when we allow the coverage achieved to fall short of
the threshold and the budget $k$ for number of seed nodes to be
overrun by a logarithmic factor, then we can achieve the required
coverage in the minimum possible propagation time, i.e., in the
time achieved by the optimal solution to MINTIME with budget
threshold $k$ and coverage threshold $\eta$ (Section~\ref{sec:mintime}, Theorem~\ref{corr2}).

\smallskip

\item On the other hand, for MINTIME under the IC model, we show that even bicriteria and
tricriteria approximations are hard. More precisely, let $R_{OPT}$ be the
optimal propagation time required for achieving a coverage $\geq \eta$
within a budget of $k$. Then we show the following (Section~\ref{sec:mintime},
Theorem~\ref{thm:bi-mt}):
there is unlikely to be a
PTIME algorithm that finds a seed set with size under the budget, which
achieves a coverage better than $(1-1/e)\eta$.
Similarly, if we limit the budget overrun factor to less than $\ln(\eta)$, then
it is unlikely that there is a PTIME algorithm that finds a seed set of size
within the overrun budget which achieves a coverage better than $(1-1/e)\eta$.
In both cases, the result holds even when we permit any amount of slack in the resulting propagation time.

\smallskip

\item The above results are bicriteria bounds, in that they allow slack
in two of the three parameters governing MINTIME problems. We also
show a tricriteria hardness result (Section~\ref{sec:mintime},
Theorem~\ref{thm:tri-mt}). Namely, if we limit the budget overrun
factor to be $\beta < \ln(\eta)$, then it is unlikely that there
is a PTIME algorithm that finds a seed set with a size within a
factor $\beta$ of the budget that achieves a coverage better than
$(1-1/e^{\beta})\eta$. Similar bounds hold if we place hard limits
on the coverage approximation and try to balance overrun in the
other parameters.

\smallskip

\item Often, the coverage function can be hard to
compute exactly. This is the case for both IC and LT models
\citep*{kempe03}. All our results are robust in that they carry over
even when only estimates of the coverage function are available.

\smallskip

\item We show the value of our approximation algorithms by
experimentally comparing their quality with that of several heuristics
proposed in other contexts, using two real
data sets. We discuss our findings in Section~\ref{sec:exp}.
\end{itemize}

\smallskip

The necessary background is given in
Section~\ref{sec:bg} while related work is discussed in
Section~\ref{sec:related}. Section~\ref{sec:dis}
concludes the paper and discusses interesting open
problems.


%% file: background.tex
Suppose we are given a social network together with the estimates
of mutual influence between individuals in the network, and
suppose that we want to push a new product in the market.  The
mining problem of \emph{influence maximization} is the following:
given such a network with influence estimates, how to select the
set of initial users so that they eventually influence the largest
number of users in the social network. This problem has received a good deal of attention in the data mining and the theoretical computer science communities in the last decade.

The first to consider the propagation of influence and the problem
of identification of influential users from a data mining
perspective are 
\cite{domingos01,domingos02}. The problem is modelled by means of
\emph{Markov random fields} and heuristics are given for choosing
the users to target. In particular, the marketing objective
function to maximize is the global expected lift in profit, that
is, intuitively, the difference between the expected profit
obtained by employing a marketing strategy and the expected profit
obtained using no marketing at all. A Markov
random field, is an undirected graphical model representing the
joint distribution over a set of random variables, where nodes are
variables, and edges represent dependencies between variables. It
is adopted in the context of influence propagation by modelling
only the final state of the network at convergence as one large
global set of interdependent random variables.

\citet{kempe03} tackle roughly the same problem
as a problem in discrete optimization. They obtain provable
approximation guarantees under various propagation models studied in
mathematical sociology, as we describe next.

A social network can be represented as a directed graph $G =
(V,E)$. Every node is in one of two states -- \emph{active} or
\emph{inactive}. Here, ``active'' may correspond to a user buying
a product or getting infected. In progressive models, it is
assumed once a node becomes active, it remains active. Influence
is assumed to propagate from nodes to their neighbors according to
a \emph{propagation model}, and a node's tendency to become active
increases monotonically as more of its neighbors become active.

In the \emph{independent cascade} (IC) model, each active neighbor
$v$ of a node $u$ has one shot at influencing $u$ and succeeds
with probability $p_{v,u}$, the probability with which $v$
influences $u$. In the \emph{linear threshold} (LT) model, each
node $u$ is influenced by each neighbor $v$ according to a weight
$ b_{v,u}$, such that the sum of incoming weights to $u$ is no
larger than $1$. Each node $u$ chooses a threshold $\theta_u$
uniformly at random from the interval $[0,1]$. If at timestamp
$t$, the total weight from the active neighbors of $u$ attains the
threshold $\theta_u$, then $u$ will become active at timestamp $t
+ 1$. In both the models, the process repeats until no new node becomes
active.

For any propagation model, the \emph{expected influence spread}
of a seed set $S$ is the expected number of nodes that eventually
get activated by initially activating the nodes $S$. We denote
this number by $\sigma_m(S)$, where $m$ stands for the underlying
propagation model. Then the \emph{influence maximization problem} is defined as follows.
Given a directed and edge-weighted social graph $G = (V, E)$, a
propagation model $m$, and a number $k \le |V|$, find a set $S
\subseteq V$, $|S| = k$, such that $\sigma_m(S)$  is maximum.

Under both the IC and LT propagation models, this problem
is shown to be \NPhard\ \citep{kempe03}. However, for both the
propagation models described above, the expected influence
spread function $\sigma_m(\cdot)$ is
\emph{monotone} and \emph{submodular}.
Monotonicity says as the set of activated nodes grows, the
likelihood of a node getting activated should not decrease.  More
precisely, a
A function $f$ from sets to reals is monotone if $f(S)
\leq f(T)$ whenever $S \subseteq T$.
A function $f$ is submodular if $f(S \cup \{w\}) - f(S) \geq f(T \cup \{w\}) -
f(T) \mbox{ whenever }  S \subseteq T$.
Submodularity intuitively
says an active node's probability of activating some inactive node $u$ does
not increase if more nodes have already attempted to activate $u$ and $u$ is hence more ``marketing-saturated''. It is also
called \emph{the law of ``diminishing returns''}.\footnote{A variant of the linear
threshold model, where a \emph{deterministic} threshold $\theta_u$
is chosen for each node, has also been studied
\citep{chen08,ben-zwi2009}. Coverage under this variant is not
submodular.}

Thanks to these two properties we can have a simple greedy algorithm
(see Algorithm~\ref{alg:greedy}) for infuence maximization which provides
an approximation guarantee. In fact, for any monotone submodular function $f$
with $f(\emptyset) = 0$, the problem of finding a set $S$ of size $k$ such
that~$f(S)$ is maximum, can be approximated to within a factor of $(1 - 1/e)$ by
the greedy algorithm~\cite{submodular}. This result
carries over to the influence maximization problem \cite{kempe03}, meaning that the seed
set we produce using Algorithm~\ref{alg:greedy} is guaranteed to have an
expected spread $(1-1/e)$ i.e., $>63\%$, of the expected spread of the optimal seed set.

The complex step of the greedy algorithm is in line 3, where we
select the node that provides the largest marginal gain $\sigma_m(S \cup \{v\}) - \sigma_m(S)$
with respect to the expected spread of the current seed set $S$. Computing
the expected spread given a seed set is \SPhard\
under both the IC model~\citep{ChenWW10} and the LT
model~\citep{ChenWW10b}. In their paper, Kempe et al. run Monte Carlo (MC)
simulations of the propagation model for sufficiently many times
(the authors report $10,000$ trials) to obtain an accurate
estimate of the expected spread, resulting in a very long
computation time. In particular, they show
that for any $\phi > 0$, there is a $\delta  > 0$ such that by
using $(1 + \delta)$-approximate values of the expected spread, we can
obtain a $(1-1/e-\phi)$-approximation for the influence
maximization problem.

\begin{algorithm}[t!]
\caption{Greedy MAXINF} \label{alg:greedy}
\begin{algorithmic}[1]
\REQUIRE{$G,k,\sigma_m$} \ENSURE{seed set $S$}
\STATE $S \leftarrow \emptyset$ \WHILE{$|S| < k$}
 \STATE $u \leftarrow \argmax_{w \in V \setminus S}(\sigma_m(S  \cup
\{w\}) - \sigma_m(S))$;
\STATE $S \leftarrow S \cup \{u\}$
\ENDWHILE
\end{algorithmic}
\end{algorithm}

We now define the problems we study in this paper. Let $m$ stand for any propagation model
with a submodular coverage function $\sigma_m(.)$.

\begin{problem}[MINTSS]
Let $G=(V,E)$ be a social graph.
Given a real number $\eta\leq |V|$, find a set $S \subseteq V$ of
the smallest size $|S|$,  such that the expected spread,
 denoted $\sigma_m(S)$, is no less than $\eta$.
\end{problem}
\begin{problem}[MINTIME]
Let $G=(V,E)$ be a social graph. Given an integer $k$, and a real
number $\eta \leq |V|$, find a set $S \subseteq V$, $|S| \leq k$,
and the smallest $t \in \mathbb{N}$, such that the expected spread
at time $t$, denoted $\sigma^t_m(S)$, is no less than $\eta$.
\end{problem}

The MINTSS problem is closely related to the real-valued submodular set cover (RSSC) problem,
defined as follows: given a submodular function $f: 2^\X \rightarrow \reals$
and a threshold $\eta$, find a set $S \subseteq \X$ of the least
size (or minimum cost, when elements of $\X$ are weighted) such
that $f(S) \geq \eta$. MINTSS under any propagation model such as
IC and LT, for which the coverage function is submodular is
clearly a special case of RSSC, an observation we exploit in Section~\ref{sec:minseed}.

MINTIME is closely related 
to the Robust Asymmetric $k$-center (RAKC)
problem in directed graphs, defined as follows: given a digraph $G = (V, E)$, a (possibly
empty) set of
forbidden nodes and thresholds $k$ and $\eta$, find $k$ or fewer
nodes $S$ such that they cover at least $\eta$ non-forbidden nodes
in the minimum possible radius, i.e., each of the $\eta$ nodes are
reachable from some node in $S$ in the minimum possible distance.

%% file: related.tex

While to the best of our
knowledge, MINTIME has never been studied before, some work has
been devoted to MINTSS. \cite{chen08} shows that under the LT
propagation model with fixed (and hence deterministic) thresholds,
MINTSS cannot be approximated within a factor of
$O(2^{\log^{1-\delta} n})$ unless $NP \subseteq DTIME(n^{O(\log
\log n)})$, and also gives a polynomial time algorithm for MINTSS
on trees.
Coverage under the LT model with deterministic thresholds is not
submodular.

\cite{ben-zwi2009} build upon \cite{chen08}
and develop a $O(n^{O(w)})$ algorithm for solving MINTSS exactly
under the deterministic linear threshold model, where $w$ is the
tree width of the graph. They show the problem cannot be solved in
$n^{O(\sqrt{w})}$ time unless all problems in SNP can be solved in
sub-exponential time. In this paper, we study both MINTSS and
MINTIME under the classic propagation models, under which the
coverage function is
submodular.%

A few classical cover-problems are related to the problems we
study. One such problem is Maximum Coverage (MC): given a
collection of sets $\mathcal{S}$ over a ground set $\U$ and budget
$k$, find a subcollection $\mathcal{C} \subseteq \mathcal{S}$ such
that $|\mathcal{C}| \le k$ and $|\bigcup \mathcal{C}|$ is
maximized. The problem can be approximated within a factor of
$(1-1/e)$ and it cannot be improved~\citep{feige98,khuller99}. Similar
results by \cite{khuller99} and \cite{sviridenko2004} exist  for the weighted case.

\eat{In the weighted
variant of the problem, the same $(1-1/e)$ approximation can be
achieved using a partial enumeration technique \cite{khuller99,
sviridenko2004}.
}

Another relevant problem is Partial Set Cover (PSC): given a
collection of sets $\mathcal{S}$ over the ground set $\mathcal{U}$
and a threshold $\eta$, the goal is to find a subcollection
$\mathcal{C} \subseteq \mathcal{S}$ such that $|\bigcup
\mathcal{C}|\ge\eta$ and $|\mathcal{C}|$ is minimized. While PSC
can be approximated within a factor of $\lceil \ln \eta \rceil$,
\cite{feige98} showed that it cannot be approximated within
a factor of $(1-\delta)\ln \eta$, for any fixed $\delta > 0$,
unless $NP \subseteq DTIME(n^{O(\log \log n)})$. 

Our results on MINTSS exploit its connection to the real-valued
submodular set cover (RSSC) problem. There has been substantial
work on submodular set cover (SSC) in the presence of
integer-valued submodular functions, which is a generalization of
the classical Set Cover Problem
\citep{fujito99,fujito00,feige98,slavik97,Judit2001}. Relatively
much less work has been done on real-valued SSC. For
non-decreasing real-valued submodular functions, \cite{wolsey82} has shown, among other things, that a simple
greedy algorithm yields a solution to a special case of SSC where
$\eta = f(\X)$, that is within a factor of $\ln[\eta/(\eta -
f(S_{t-1})]$ of the optimal solution, where $t$ is the number of
iterations needed by the greedy algorithm to achieve a coverage of
$\eta$ and $S_i$ denotes the greedy solution after $i$ iterations.
Unfortunately, this result by itself does \emph{not} yield an
approximation algorithm with any guaranteed bounds: in
Appendix~\ref{sec:wolsey}
we give an example to show that the
greedy solution can be arbitrarily worse than the optimal one.
Furthermore, Wolsey's analysis is restricted to the case $\eta =
f(\X)$. Along the way to establishing our results on MINTSS, we
show the greedy algorithm yields a bicriteria approximation for
real-valued SSC that extends to the general case of partial cover
with $\eta \leq f(\X)$, and where elements are weighted. \eat{
\footnote{This result holds when elements of $\X$ are weighted and
one seeks to minimize the sum of weights of the solution.}}

Our results on MINTIME leverage its connection to the robust
asymmetric $k$-center problem (RAKC). It has been shown that,
while asymmetric $k$-center problem can be approximated within a
factor of $O(\log^* n)$ \citep{sundar1998},
RAKC cannot be approximated within any factor
unless P = NP \citep{gortz2006}.


%% file: minseed.tex
\subsection{A Bicriteria Approximation}
\label{sec:bc-ms} Our main result of this section is that a simple
greedy algorithm, Algorithm $\MINSEEDSET$, yields a bicriteria
approximation to (weighted) MINTSS, for any propagation model
whose coverage function is monotone and submodular.



\begin{algorithm}[t!]
\caption{ $\MINSEEDSET$} \label{alg:mintss}
\begin{algorithmic}[1]
\REQUIRE{$G,\eta,\epsilon, \sigma_m$}
\ENSURE{seed set $S$}
\STATE $S \leftarrow \emptyset$
\WHILE{$\sigma_m(S) < \eta - \epsilon$}
	\STATE $u \leftarrow \argmax_{w \in V \setminus S}(\frac{\min(\sigma_m(S  \cup
\{w\}), \eta) - \sigma_m(S)}{c(w)})$;
\STATE $S \leftarrow S \cup \{u\}$
\ENDWHILE
\end{algorithmic}
\end{algorithm}

In order to prove the results in the most general setting, we
consider digraphs $G=(V,E)$ which have non-negative node weights:
we are given a cost function $c: V \rightarrow \reals^+$ in
addition to the coverage threshold $\eta$, and need to find a seed
set $S$ such that $\sigma_m(S) \ge \eta$ and $c(S) = \sum_{x\in S}
c(x)$ is minimum. Clearly, this generalizes the unweighted case.








\begin{theorem}\label{thm:sms_cost-ms}
Let $G = (V,E)$ be a social graph, with node weights given by
$c: V \rightarrow \reals^+$. Let $m$ be any propagation model
whose coverage function $\sigma_m(.)$ is monotone and submodular.
Let $\S^*$ be a seed set of minimum cost such that
$\sigma_m(\S^*) \ge \eta$. Let $\epsilon > 0$ be any shortfall and let $S$ be the
greedy solution with chosen threshold $\eta-\epsilon$.
Then, $c(S) \le c(S^*) \cdot (1 + \ln (\eta/\epsilon))$.
\end{theorem}

In the rest of this section, we prove this result. We first
observe that every instance of MINTSS where the coverage function
$\sigma_m(.)$ is monotone and submodular is an instance of RSSC.
Thus, it suffices to prove Theorem~\ref{thm:sms_cost-ms} for RSSC,
for which we adapt a bicriterion approximation technique by
\cite{slavik97}.

Let $\X = \{x_1, x_2, ..., x_m\}$ be a ground set,
$c: \X \ra \reals^+$ be a cost function,
$f:2^{\X} \ra \reals$ a non-negative monotone
submodular function and $\eta$ a given threshold.
Apply the greedy algorithm above
to this instance of RSSC.
Let $S_i$ be the (partial) solution obtained by the greedy algorithm
after $i$ iterations. Let $t$ be the smallest number such that
$f(S_t) \ge \eta$. We define $g(S) = \min (f(S), \eta)$. Clearly, $g$ is also
monotone and submodular. In each iteration, the greedy algorithm picks an
element which provides the maximum marginal gain per unit cost (w.r.t. $g$),
i.e., it picks an element $x$ for which $\frac{g(S \cup \{x\}) -
g(S)}{c(x)}$ is positive and is maximum.

Let $c(S^*)=\kappa$ and
define $\eta_i = \eta - g(S_i)$, i.e., the shortfall in
coverage after $i$ iterations of the greedy algorithm.


\begin{lemma}
\label{lem:greedy}
At the end of iteration $i$, there is an element $x \in \X \setminus \S_i$:
$\frac{g(S_i \cup \{x\}) - g(\S_i)}{c(x)} \ge \frac{\eta_i}{\kappa}$.
\end{lemma}

\spara{Proof.} Let $\S_i^* = \S^* - \S_i$. Let $\S_i^* = \{y_1,
..., y_t\}$ and $c(S_i^*) = \kappa_i$. Suppose $\forall x \in \X
\setminus \S_i: \frac{g(\S_i \cup \{x\}) - g(\S_i)}{c(x)} <
\frac{\eta_i}{\kappa}$. Consider adding the elements in $\S_i^*$
to $\S_i$ one by one. Clearly, at any step $j \le t$, we have by submodularity that
\begin{align*}
g(\S_i \cup \{y_1,..., y_{j}\}) &- g(\S_i \cup \{y_1,...,
y_{j-1}\}) \\
&\le g(\S_i \cup \{y_{j}\}) - g(\S_i) < {c(y_{j})} \cdot \frac{\eta_{i}}{\kappa}
\end{align*}
Iterating over all $j$, this yields $g(\S_i \cup \{y_1,..., y_{j}\}) - g(\S_i) < \frac{\eta_i}{\kappa} \cdot (c(y_1) + ... + c(y_j))$
resulting in $g(\S_i \cup \{y_1,..., y_{t}\}) < g(S_i) + \frac{\eta_i}{\kappa} \cdot \sum_{1 \le j \le t} c(y_j) \le \eta$
which is a contradiction since the left hand side is no less than the optimal coverage. \qed




\medskip
\noindent
\textbf{Proof of Theorem~\ref{thm:sms_cost-ms}}:

It follows from Lemma~\ref{lem:greedy} that $\eta_i \le \eta_{i-1} (1 - c_i/\kappa)$ where $c_i$ is
the cost of the element added in iteration $i$. Using the well known
inequality $(1+z) \le e^z, \forall z$, we get $\eta_i \le \eta_{i-1} \cdot
e^{-c_i/k}$. Expanding, $\eta_i \le \eta \cdot e^{- \frac{1}{k} \cdot
\sum_{i} c_i}$.
Let the algorithm take $l$ iterations to achieve coverage $g(S_l) \ge \eta -
\epsilon$ such that $g(S_{l-1}) < \eta - \epsilon$. At any step,
$g(S_{i+1}) - g(S_i) \le \eta_i$. Thus, $c_i
\le \kappa$, and in particular, the cost of the last element picked can be at most $\kappa$. So,
$c(S_l) \le \kappa + c(S_{l-1})$. $g(S_{l-1}) < \eta - \epsilon$ implies $\eta_{l-1} > \epsilon$. Hence, we have
$\eta e^{- \frac{1}{\kappa} c(S_{l-1})} > \epsilon$ which implies
$c(S_{l-1}) < \kappa \ln (\eta/\epsilon)$.
Thus, $c(S_l) \le \kappa (1 + \ln (\eta/\epsilon))$.
\qed

Using a similar analysis, it can be shown that when the costs are
uniform, the approximation factor can be improved to $\lceil \ln
(\eta/\epsilon) \rceil$.


For propagation models like IC and LT,
computing the coverage
$\sigma_m(S)$ exactly is \SPhard\ \citep{ChenWW10,ChenWW10b} and
thus we must settle for estimates. To address this, we ``lift'' the
above theorem to the case where only estimates of the function
$f(.)$ are available. We can show:

\begin{theorem}
\label{lem:monte-carlo} For any $\phi > 0$, there exists a $\delta
\in (0, 1)$ such that using ($1 - \delta$)-approximate values for
the coverage function $\sigma_m(\cdot)$, the greedy algorithm
approximates MINTSS under IC and LT models within a factor of $(1
+ \phi) \cdot (1 + \ln (\eta/\epsilon))$.
\end{theorem}

\noindent
\textbf{Proof.}
The proof involves a more careful analysis of how error propagates in
the greedy algorithm if, because of errors, the greedy algorithm picks
the wrong point.

  Here, we give the proof for the unit cost version only. Consider any
  monotone, submodular function $f(\cdot)$. Thus, in the statement of
  theorem, $\sigma_m(\cdot) = f(\cdot)$. Let $f'(\cdot)$ be its
  approximated value. In any iteration, the (standard) greedy
  algorithm picks an element which provides maximum marginal gain.
  Let $S_i$ be the set formed after iteration $i$.


  As we did in Lemma \ref{lem:greedy}, it is straightforward to show
  that there must exists an element $x \in \X \setminus S_i$ such that
  $f(S_i \cup \{x\}) - f(S_i) \ge \eta_i/k$ where $\eta_i = \eta -
  f'(S_i)$. Without loss of generality, let $x$ be the element which
  provides the maximum marginal gain. Suppose that due to the error in
  computing $f(.)$, some other element $y$ is picked instead. Then,
  \begin{align*}
    (1-\delta)f(S_i \cup \{x\}) \le f'(S_i \cup \{x\}) \le f'(S_i \cup
    \{y\})
  \end{align*}

  Moreover, $f'(S_i) \le f(S_i)$. Thus,
  \begin{align*}
    &\frac{\eta_i}{k} \le f(S_i \cup \{x\}) - f(S_i) \le \frac{f'(S_i
      \cup
      \{y\})}{1-\delta} - f'(S_i) \\
    \implies &\frac{\eta_i}{k} \le \frac{\eta - \eta_{i+1}}{1-\delta}
    - \eta
    +\eta_i  \\
    \implies & \eta_{i+1} \le \eta_i \cdot (1-\delta) \cdot \left( 1 -
      \frac{1}{k} \right) + \delta \cdot \eta \\
    \implies & \eta_{i+1} \le \eta \cdot (1-\delta)^{i+1} \cdot \left(
      1 - \frac{1}{k} \right)^{i+1} \\
	  &+ \delta \cdot \eta \cdot \left(
      \frac{1 - (1 - \delta)^{i+1} (1 - 1/k)^{i+1}}{1 - (1 - \delta)
        (1 - 1/k)} \right)
  \end{align*}

  Let $\delta' = \delta / (1 - (1 - \delta)(1 - 1/k))$. Let the greedy
  algorithm takes $l$ iterations. Then,
  \begin{align*}
    \eta_{l} &\le \eta \cdot (1-\delta)^{l} \cdot \left( 1 -
      \frac{1}{k} \right)^{l} \\
	  &+ \delta' \cdot \eta \cdot \left( 1 -
      (1-\delta)^{l} \cdot \left( 1 -
        \frac{1}{k} \right)^{l} \right) \\
    &= \eta \cdot (1-\delta)^{l} \cdot \left( 1 - \frac{1}{k}
    \right)^{l} (1 - \delta') + \delta' \cdot \eta
  \end{align*}

  Using $(1-\delta)^l \le 1$ and $(1-1/k)^l \le e^{-l/k}$,
  \begin{align*}
    \eta_{l} \le \eta e^{-l/k} (1 - \delta') + \delta' \cdot \eta
  \end{align*}

  The algorithm stops when $\eta_l \le \epsilon$. The maximum number
  of iterations needed to ensure this are
  \begin{align*}
    l \le k \left( 1 + \ln \frac{\eta (1-\delta')}{\epsilon (1 -
        \delta' \eta/\epsilon)} \right)
  \end{align*}

  Let $x = \eta/\epsilon$. To prove the lemma, we need to prove that
  for any $\phi > 0$, there exists $\delta \in [0, 1)$ such that
  \begin{align*}
    &x^{1+\phi} = x \frac{1-\delta'}{1 - \delta' x} \implies \delta' =
    \frac{x^\phi - 1}{x^{1+\phi} - 1}
  \end{align*}

  Clearly, for any $\phi \ge 0$, $\delta' \in [0, 1)$. Hence,
  \begin{align*}
    &0 \le \delta < 1 - (1 - \delta)(1 - 1/k) \\
    \Longleftrightarrow & 0 \le \delta < 1
  \end{align*}

  This completes the proof for unit cost case. Using the slight
  modification in the greedy algorithm (as we did in proving theorem
  1), the same result can be obtained for weighted version. \qed

\subsection{An Inapproximability Result}



Recall that every instance of MINTSS where the coverage function
is monotone and submodular is an instance of RSSC. Consider the
unweighted version of the RSSC problem. Let $S^*$ denote an
optimal solution and let $\OPT = |S^*|$.





\begin{theorem}\label{thm:inapprox3}
For any fixed $\delta > 0$, there does not exist a PTIME algorithm for
RSSC that guarantees a solution  $\S: |\S| \le \OPT
(1-\delta) \ln(\eta/\epsilon)$, and $f(\S) \ge \eta - \epsilon$ for any
$\epsilon > 0$ unless $NP \subseteq DTIME(n^{O(\log \log n)})$.
\end{theorem}

\noindent
\textbf{Proof.}
\textbf{Case 1: $\epsilon \ge 1$.\ }
Suppose there exists an algorithm $\A$ that finds a solution
$S$ of size $\le \OPT (1-\delta)\ln(\eta/\epsilon)$ such that $f(S) \ge
\eta - \epsilon$ for any $\epsilon \ge 1$. Consider an arbitrary
instance $\mathcal{I} = \langle \mathcal{U}, \mathcal{S}, \eta \rangle$
of PSC, which is a special case of RSSC.  Apply the algorithm $\A$ to $\I$.
It outputs a collection of sets $\C_1: |\C_1| \le
\OPT (1-\delta)\ln(\eta/\epsilon)$ that covers $\ge \eta - \epsilon$
elements in $\mathcal{U}$.

Create a new instance $\mathcal{J} = \langle \mathcal{U}',
\mathcal{S}', \eta' \rangle$ of PSC as follows. Let $T = \bigcup \C_1$
be the set of elements of $\mathcal{U}$ covered by $\C_1$.
Define $\mathcal{S}' = \{S \setminus T \mid
\S\in \mathcal{S} \setminus \C_1\}$, $\mathcal{U}' =
\mathcal{U} \setminus T$ and $\eta' = \epsilon$.
Set the new shortfall $\epsilon' = 1$.
Apply the algorithm $\A$ to $\mathcal{J}$. It will output another collection of sets $\C_2: |\C_2| \le \OPT
(1-\delta)\ln \epsilon$ which covers
$\ge \epsilon - 1$ elements in $\mathcal{U}'$.\footnote{If $\epsilon = 1$, $\A$ outputs
an empty collection.}
Let $\C = \C_1 \cup \C_2$.
The number of elements covered by $\C$ is $\ge \eta - \epsilon +
\epsilon - 1 = \eta - 1$.
Clearly, $|\C| = |\C_1| + |\C_2| \le \OPT (1-\delta)\ln(\eta/\epsilon)
+ \OPT (1-\delta)\ln(\epsilon)  =  \OPT (1-\delta)\ln(\eta)$.
Thus, we have a solution for PSC with the
approximation factor of $(1-\delta)\ln(\eta)$, which is not
possible unless $NP \subseteq DTIME(n^{O(\log \log n)})$ \citep{feige98}. This proves Case 1.

\textbf{Case 2: $\epsilon < 1$.\ }
Assume an arbitrary instance $\mathcal{I}$ of RSSC with monotone
submodular function $\f: 2^{\X} \rightarrow \mathbb{R}$. Let
$\eta'$ be the coverage threshold and $\epsilon' \ge 1$ be any
given shortfall. We now construct another instance $\mathcal{J}$
of RSSC as follows: Set the coverage function $\g(S) = \f(S)/x$,
coverage threshold $\eta = \eta'/x$ and shortfall $\epsilon =
\epsilon'/x$. Choose any value of $x > 1$ such that $\epsilon =
\epsilon'/x < 1$. We now show that if a solution is a
$(1-\delta)\ln(\eta/\epsilon)$-approximation to the optimal
solution for $\mathcal{J}$ then it is a
$(1-\delta)\ln(\eta'/\epsilon')$-approximation to the optimal
solution for $\mathcal{I}$. Clearly, the optimal solution for both
the instances are identical, so $\OPT_\mathcal{I} =
\OPT_\mathcal{J}$.\footnote{Here, $\OPT_\I$ and $\OPT_\J$
represent the size of the optimal solution for instances $\I$ and
$\J$ respectively.} Suppose there exists an algorithm for RSSC
when the shortfall is $\epsilon \in (0, 1)$, that guarantees a
solution $\S: |\S| \le \OPT (1-\delta) \ln(\eta/\epsilon)$ and
$\f(\S) \ge \eta - \epsilon$. Apply this algorithm to instance
$\mathcal{J}$ to obtain a solution $\S_\mathcal{J}$. We have:
$\g(\S_\mathcal{J}) \ge \eta - \epsilon = (\eta' - \epsilon')/x$.
It implies $\f(\S_\mathcal{J}) = x \cdot \g(\S_\mathcal{J}) \ge
\eta' - \epsilon'$. Moreover, $|\S_\mathcal{J}| \le
\OPT_\mathcal{J} (1 - \delta) \ln (\eta/\epsilon)$, implying
$|\S_\mathcal{J}| \le \OPT_\mathcal{I} (1 - \delta) \ln
(\eta'/\epsilon')$. Thus we have the solution $S_\mathcal{J}$ for
instance $\mathcal{I}$ whose size is $\le \OPT_\mathcal{I} (1 -
\delta) \ln (\eta'/\epsilon')$. The theorem follows. \qed

In view of this generic result, we conjecture that improving the
approximation factor for MINTSS to $(1 - \delta) \ln
(\eta/\epsilon)$ for IC and LT is likely to be hard.

%% file: mintime.tex
%
%
%

In this section, we study MINTIME under the IC model. Denote by
$\sigma_m^R(S)$ the expected number of nodes activated under model
$m$ within time $R$, and let $\eta$ be the desired coverage and
$k$ be the desired budget. Let $R_\OPT$ denote the optimal
propagation time under these budget and coverage constraints. Our first
result says that  efficient approximation algorithms are unlikely
to exist under two scenarios: (i) when we allow a coverage
shortfall of less than $\eta/e$ and (ii) when we allow a budget
overrun less than $\ln \eta$. In the former scenario, we have a
strict budget threshold and in the latter we have a strict
coverage threshold. In both cases, we  allow any amount of slack
in propagation time.

\begin{theorem}\label{thm:bi-mt}
Unless $NP \subseteq DTIME(n^{O(\log \log n)})$, there does not exist a PTIME
algorithm for MINTIME that guarantees (for
any $\alpha \ge 1$):
\begin{enumerate}
\item a ($\alpha, \gamma$)-approximation, such that $|\S| \le k$,
$R = \alpha \cdot R_{\OPT}$ and $\sigma_m^R(\S) \ge \gamma \cdot \eta$ where $\gamma =
(1 - 1/e + \delta)$ for any fixed $\delta > 0$; or

\item a ($\alpha, \beta$)-approximation, such that $|\S| \le \beta
\cdot k$, $R = \alpha \cdot R_{\OPT}$ and $\sigma_m^R(\S) \ge \eta$ where
$\beta = (1 - \delta) \ln \eta$ for any fixed $\delta > 0$.

\end{enumerate}
\end{theorem}

Our second theorem says efficient approximation algorithms are
unlikely to exist under more liberal scenarios than those given above:
(i) when for a given budget overrun factor $\beta < \eta$, the
fraction of the coverage we want to achieve is more than $1 -
1/e^{\beta}$ and (ii) when for a given fraction $\gamma \in (0, 1-1/\eta]$ of the coverage we want to achieve, the budget overrun factor we allow is
less than $\ln (1/(1-\gamma))$. As before, we allow any amount of slack
in propagation time.

\begin{theorem}\label{thm:tri-mt}
Unless $NP \subseteq DTIME(n^{O(\log \log n)})$ there does not
exist a PTIME
algorithm for MINTIME that guarantees ($\alpha, \beta, \gamma$)-approximation
factor (for any $\alpha \ge 1$) such that $|\S| \le \beta \cdot k$, $R = \alpha \cdot
R_{\OPT}$ and $\sigma_m^R(\S) \ge \gamma \cdot \eta$ where
\begin{enumerate}
\item $\beta \in [1, \ln \eta)$ and $\gamma = 1 - 1/e^\beta + \delta$ for any
fixed $\delta > 0$; or

\item $\gamma \in \left( 0,  1 - \frac{1}{\eta} \right]$ and
$\beta = (1 - \delta) \ln \left( \frac{1}{1-\gamma} \right)$ for
any fixed $\delta > 0$.
\end{enumerate}
\end{theorem}


Finally, on the positive side, we show that when a coverage shortfall of
$\epsilon > 0$ is allowed and a budget boost of $(1+\ln(\eta/\epsilon))$ is
allowed, we can in PTIME find a solution which achieves the relaxed coverage
under the relaxed budget in optimal propagation time. More precisely, we have:

\begin{theorem} \label{corr2}
Let the chosen coverage threshold be $\eta - \epsilon$, for $\epsilon > 0$ and
chosen budget threshold be $k(1+\ln(\eta/\epsilon))$.
If the coverage function $\sigma_m^R(\cdot)$ can be computed exactly, then
there is a greedy algorithm that approximates the MINTIME problem within a
($\alpha, \beta, \gamma)$ factor where
$\alpha = 1$, $\beta = 1 + \ln(\eta/\epsilon)$ and $\gamma = 1 -
\epsilon/\eta$ for any $\epsilon > 0$.
Furthermore, for every $\phi > 0$,
there is a $\delta > 0$ such that by using a $(1-\delta)$-approximate
values for the coverage function $\sigma_m^R(\cdot)$, the greedy algorithm
approximates the MINTIME problem within a ($\alpha, \beta, \gamma)$
factor where $\alpha = 1$, $\beta = (1+\phi)(1+\ln (\eta/\epsilon))$ and
$\gamma = 1-\epsilon/\eta$.
\end{theorem}



\subsection{Inapproximability Proofs}
\label{subsec:akc}
\input{akc1}

\subsection{A Tri-criteria Approximation}
\label{sec:tri-mt}  We now consider upper bounds for MINTIME. It
is interesting to ask what happens when either the budget overrun or
the coverage shortfall is increased. We show that under these
conditions, a greedy strategy combined with linear search yields a
solution with optimal propagation time. This proves
Theorem~\ref{corr2}.

Algorithm $\MINSEEDSET$ computes a small seed set $S$ that achieves
coverage $\sigma_m(S) = \eta-\epsilon$. Recall that  $\sigma_m^R(S)$
denotes the coverage of $S$ under propagation model $m$ within $R$ time
steps. It is easy to see that $\MINSEEDSET$ can be adapted to instead
compute a seed set that yields coverage $\eta - \epsilon$ within $R$
time steps: we call this algorithm $\MINSEEDSET$$^R$.

Given such an algorithm, a simple linear search over $R = 0 \ldots n-1$
yields the bounds specified in Theorem~\ref{corr2}, after  setting
coverage threshold as $\eta - \epsilon$
and the chosen budget threshold as $\mathit{budg} = k(1+\ln(\eta/\epsilon))$. The approximation
factors in the theorem follow from Theorem~\ref{thm:sms_cost-ms} and
Lemma~\ref{lem:monte-carlo}. These bounds continue to hold if we can only provide estimates for the coverage function (rather than computing it exactly) and also extend to weighted nodes.

We conclude this section by noting that the algorithm above can be
naturally adapted to the RAKC problem. The bounds in
Theorem~\ref{corr2} apply to RAKC as well, since MINTIME under
IC generalizes RAKC.


%% file: akc1.tex
 We next prove
Theorems~\ref{thm:bi-mt} and~\ref{thm:tri-mt}. We first show that
MINTIME under the IC model generalizes the RAKC problem. In a
digraph $G = (V,E)$ and sets of nodes $S, T \subset V$, say that
\emph{$R$-covers} $T$ if for every $y\in T$, there is a $x\in S$
such that there is a path of length $\le R$ from $x$ to $y$. Given
an instance of RAKC, create an instance of MINTIME by labeling
each arc in the digraph with a probability $1$. Now, it is easy to
see that for any set of nodes $S$ and any $0 \leq R\leq n-1$, $S$
$R$-covers a set of nodes $T$ iff activating the seed nodes $S$
will result in the set of nodes $T$ being activated within $R$
time steps. Notice that since all the arcs are labeled with
probability $1$, all influence attempts are successful by
construction. It follows that RAKC is a special case of MINTIME
under the IC model.

The tricriteria inapproximability results of
Theorem~\ref{thm:tri-mt} subsume the bicriteria inapproximability
results of Theorem~\ref{thm:bi-mt}. Still, in our presentation, we
find it convenient to develop the proofs first for bicriteria.
Since we showed that MINTIME under IC generalizes RAKC, it
suffices to prove the theorems in the context of RAKC. It is worth
pointing out \cite{gortz2006} proved that it is
hard to approximate $\RAKC$ within any factor unless $P = NP$.
Their proof only applies to (the standard) unicriterion
approximation.

For a set of nodes $S$ in a digraph we denote by $f^R(S)$ the number of
nodes that are $R$-covered by $S$. Recall the problems MC and PSC (see
Section~\ref{sec:related}).

\smallskip

\noindent
{\bf Proof of Theorem~\ref{thm:bi-mt}}:
It suffices to prove the theorem for RAKC.
For claim 1, we reduce Maximum Coverage (MC) to RAKC and for claim
2, we reduce PSC to RAKC. The reduction is similar and is as follows:
Consider an instance of the decision version of MC (equivalently PSC) $\mathcal{I}=\langle
\mathcal{U}, \mathcal{S}, k, \eta \rangle$, where we ask whether there exists
a subcollection $\C \subseteq \mathcal{S}$ of size $\leq k$ such that
$|\bigcup_{S\in\mathcal{S}} S| \geq \eta$.
Construct an instance $\mathcal{J}=\langle \mathcal{G}, k', \eta' \rangle$
of RAKC as follows:  the graph
$\mathcal{G}$ consists of  two
classes of nodes -- $A$ and $B$. For each $S\in \mathcal{S}$, create a
class A node $v_S$ and for each $u\in U$, create a class B node $v_u$.
There is a directed edge $(v_S, v_u)$ of unit length iff $u\in S$.
Notice, a set of nodes $S$ in $\mathcal{G}$ $R$-covers
another non-empty set of nodes iff $S$ $1$-covers the latter set. Moreover,
$x$ sets in $\mathcal{S}$ cover $y$ elements in
$\mathcal{U}$ iff $\mathcal{G}$ has a set of $x$ nodes which $1$-covers
$y + x$ nodes. The only-if direction is trivial. For the if direction,
the only way $x$ nodes can $1$-covers $y+x$ nodes in $\mathcal{G}$
is when the $x$ nodes are from class A.

Next, we prove the first claim. Set $k'=k$ and $\eta'=\eta+k$.
Assume there exists a PTIME ($\alpha$, $\gamma$)-approximation algorithm
$\A$ for RAKC such that $\f^R(\S) \ge (1-1/e+\delta) \cdot (\eta')$ for
any fixed $\delta > 0$, for some $R \leq \alpha R_{OPT}$.
Apply algorithm $\A$ to the instance $\mathcal{J}$. Notice, for our
instance, $R_{OPT} = 1$. The
coverage by the output seed set $\S$ will be $\f^R(\S) \ge (1-1/e+\delta) \cdot
(\eta+k)$ nodes, for some $R \leq \alpha\cdot 1$,
implying that the number of class B nodes covered is
$\ge (1 - 1/e + \delta) \cdot (\eta + k) - k$
$= ( 1 - 1/e + \delta - (1/e - \delta)
k/\eta)\eta$.
Thus the algorithm approximates MC within
a factor of $\left( 1 - \frac{1}{e} + \delta - \left( \frac{1}{e} - \delta \right)
\frac{k}{\eta} \right)$. Let $\delta' = \delta - \left( \frac{1}{e} - \delta \right)
\frac{k}{\eta}$.
If we show $\delta' > 0$, we are done, since MC cannot be
approximated within a factor of $(1-1/e+\delta')$ for any $\delta' > 0$
unless $NP \subseteq DTIME(n^{O(\log \log n)})$ \citep{feige98,khuller99}. Clearly, $\delta'$ is not always positive. However, for a given $\delta$
and $k$, $\delta'$ is an increasing function of $\eta$ and reaches $\delta$ in the
limit. Hence there is a value $\eta_0: \forall \eta\geq \eta_0$, $\delta' > 0$.
That is, there are infinitely many instances of PSC for which $\A$ is a
$(1-1/e+\delta')$-approximation algorithm, where $\delta'>0$, which proves the first claim.

Next, we prove the second claim. Set $k' = k$ and $\eta' = \eta + x$.
The value of $x$ will be decided later. Assume there exists
a PTIME ($\alpha$, $\beta$)-approximation algorithm $\A$ for RAKC
where $\beta = (1-\delta) \ln (\eta')$ for any fixed
$\delta > 0$. Apply the algorithm to $\mathcal{J}$. It gives a solution $S$
such that $|S| \le k \cdot (1-\delta) \ln (\eta + x)$ that covers $\geq \eta+x$ nodes.
A difficulty arises here since $\delta$ can be arbitrarily close to $1$
making $k \cdot (1-\delta) \ln (\eta + x)$ arbitrarily small, for any given $\eta$
and $k$. However, as we argued in the proof of claim 1, for sufficiently large
$\eta$, we can always find an $x$: $k \leq x \leq k \cdot (1-\delta) \ln (\eta + x)$.
That is, on infinitely many instances of PSC, algorithm $\A$ finds a set of $|S|$ class A nodes
which $R$-covers $\eta+x$ nodes, for some $R \leq \alpha\cdot 1$.
Without loss of generality, we can assume $x \le \eta$. Choose the smallest value of $x$ such that the solution
$S$ covers $\geq \eta$ class B nodes. This implies the number of class A nodes covered
is $\leq x$ and so $|S| \leq x$.
Thus,  on all such instances, algorithm $\A$ gives a solution $S$ of size $\leq x$: $k \le x \le k
\cdot (1-\delta)
\ln (\eta + x)$ that covers $\geq \eta$ nodes.
If we show that the upper bound is equal to $k \cdot (1-\delta') \ln \eta$
for some
$\delta' > 0$, we are done, since PSC cannot be approximated within a factor
of $(1-\delta') \ln \eta$ unless $NP \subseteq DTIME(n^{O(\log \log
n)})$ \citep{feige98}.

Let $(1-\delta') \ln \eta = (1-\delta) \ln (\eta + x)$, which yields
$\delta' = 1 - (1-\delta) \frac{\ln (\eta + x)}{\ln \eta}$. It is
easy to see that by choosing sufficiently large $\eta$, we can make
the gap between $\delta$ and $\delta'$ arbitrarily small and thus
can always ensure $\delta' > 0$ on infinitely many instances of PSC,
on each of which algorithm $\A$ will serve as an $(1-\delta')\ln \eta$-approximation
algorithm proving claim $2$. \qed

\smallskip

Note, in the proofs of both claim $1$ and $2$ in the above theorem, by choosing $\eta$ sufficiently
large, we can always ensure for any given $k$ and $\delta > 0$, the corresponding
$\delta'$ is always greater than $0$. To prove the tricriteria hardness results,
we need the
following lemma.  

\begin{lemma} \label{lem:set-tri}
In the MC (or PSC) problem, let $k$ be the minimum number of sets needed
to cover $\ge \eta$ elements. Then, unless $NP \subseteq
DTIME(n^{O(\log \log n)})$, there does not exist a PTIME algorithm that
is guaranteed to select $\beta k$ sets covering $\ge \gamma \eta$ elements
where
\begin{enumerate}
\item $\beta \in [1, \ln \eta)$ and $\gamma > 1 - 1/e^\beta$; or


\item $\gamma \in \left( 0, 1 - \frac{1}{\eta} \right]$ and
$\beta = (1 - \delta) \ln \left( \frac{1}{1 -
\gamma} \right)$ for any fixed $\delta > 0$.
\end{enumerate}
\end{lemma}

\noindent Lemma~\ref{lem:set-tri} is proved in Appendix~\ref{sec:proof:lem:set-tri}. We are ready to prove Theorem~\ref{thm:tri-mt}.

\smallskip
\noindent
{\bf Proof of Theorem~\ref{thm:tri-mt}}:
Again, it suffices to prove the theorem for RAKC.
For claim 1, we reduce MC to RAKC and for claim
2, we reduce PSC to RAKC. The reduction is the same as
in the proof of Theorem~\ref{thm:bi-mt} and we skip the details here.
Below, we refer to instances $\I$ and $\J$ as in that proof.

We first prove claim 1. Given any $\beta$, set $k'=k$ and
$\eta'=\eta+ \beta k$.  Assume there exists a PTIME ($\alpha$, $\beta$,
$\gamma$)-approximation algorithm $\A$ for RAKC which approximates the
problem within the factors as mentioned in claim 1.
Apply algorithm $\A$ to the instance $\mathcal{J}$. The
coverage by the output seed set $\S$ will be $\f^R(\S) \ge (1-1/e^\beta+\delta) \cdot
(\eta+ \beta k)$ nodes, implying the number of class B nodes covered is
$\ge (1 - 1/e^\beta + \delta) \cdot (\eta + \beta k) - \beta k$
$= ( 1 - 1/e^\beta + \delta - (1/e^\beta - \delta)
\beta k/\eta)\eta$.
Thus the algorithm approximates MC within
a factor of $\left( 1 - \frac{1}{e^\beta} + \delta - \left(
\frac{1}{e^\beta} - \delta \right)
\frac{\beta k}{\eta} \right)$.

If we show $\delta - \left( \frac{1}{e^\beta} - \delta \right)
\frac{\beta k}{\eta} > 0$, then the claim follows, since MC cannot
be approximated within a factor of $(1-1/e^\beta+\delta')$ for any
$\delta' > 0$ unless $NP \subseteq DTIME(n^{O(\log \log n)})$, by
Lemma~\ref{lem:set-tri}. Let $\delta' = \delta - \left(
\frac{1}{e^\beta} - \delta \right) \frac{\beta k}{\eta}$. For any
$\beta \in [1, \ln \eta)$, $\delta'$ is an increasing function of
$\eta$ which approaches $\delta$ in the limit. Thus, given any
fixed $\delta > 0$, there must exist some $\eta_o$ such that for
any $\eta \ge \eta_o$, $\delta' > 0$. This proves the first claim
(by an argument similar to that in Theorem~\ref{thm:bi-mt}).

Next, we prove the second claim. Set $k' = k$ and $\eta' = \eta + x$.
The value of $x$ will be decided later. Assume that there exists a PTIME ($\alpha$, $\beta$,
$\gamma$)-approximation algorithm $\A$ for RAKC where the factors
$\alpha$, $\beta$ and $\gamma$ satisfy the conditions as mentioned
in claim 2.  Apply the algorithm to instance $\mathcal{J}$. For any
$\gamma_j \in (0, 1 - 1/(\eta+x)]$, it gives a solution of size $\le k \cdot
(1-\delta) \ln (1/(1-\gamma_j))$ that covers $\gamma_j \cdot (\eta + x)$
nodes. There can be $|\mathcal{S}|$ possible choices of $x$. Pick the
smallest $x$ such that number of nodes covered in class B is at least
$\gamma_j \eta$, implying that the number of nodes picked from class A is
$\gamma_j x$. Thus, $\gamma_j x \le k \cdot (1-\delta) \ln
(1/(1-\gamma_j))$. The existence of $x$ satisfying this inequality can be
established as done for claim 2 in Theorem~\ref{thm:bi-mt}.

Thus, algorithm $\A$ gives the solution instance $\I$ of size $\le
k \cdot (1-\delta) \ln (1/(1-\gamma_j))$ that covers $\gamma_j \eta$
elements in $\U$ where $\gamma_j \in (0, 1-1/(\eta+x)]$. If we show that
for any given $\delta > 0$ and $\gamma_j$ in the range, there exists
some $\delta' > 0$ and $\gamma_i \in (0, 1 - 1/\eta]$ such that $\gamma_i\eta \ge
\gamma_j(\eta+x)$ and $(1-\delta') \ln (1/(1-\gamma_i)) =
  (1-\delta) \ln (1/(1-\gamma_j))$, then the claim follows. Let $Z =
  \left( \ln \frac{1}{1-\gamma_j} \right) / \left( \ln
  \frac{1}{1-\gamma_i} \right) $, then
  $\delta' = 1 - (1-\delta) Z$.

Whenever $\gamma_j \leq 1-1/\eta$, we can always choose $\gamma_i\ge \gamma_j$
such that $\delta' > 0$. The non-trivial case is when $\gamma_j \in (1-1/\eta, 1-1/(\eta+x)]$.
In this case, by choosing a large enough $\eta$, we can make $Z$ arbitrarily close to
$1$ and make $\delta' > 0$. In other words, there exists some $\eta_0$:
for all $\eta \geq \eta_0$, $\delta' > 0$, and by an argument similar to that for
claim 2 in Theorem~\ref{thm:bi-mt}, the claim follows.
\qed

\smallskip



%% file: exp.tex
We conducted several experiments to assess the value
of the approximation algorithms by comparing their quality against
that achieved by several well-known heuristics, as well as against the
state-of-the-art methods developed for MAXINF that we adapt in order to
deal with MINTSS and MINTIME. In particular, the goals of experimental evaluation are two-fold. First, we
have previously established from theoretical analysis that the Greedy
algorithm (\textsc{Greedy-Mintss} for MINTSS and
\textsc{Greedy-Mintss$^{R}$} for MINTIME) provides the best possible
solution that can be obtained in PTIME, which we would like to validate empirically.
Second, we study the gap between the solutions obtained from various
heuristics against the Greedy algorithm, the upper bound, in terms of quality.

In what follows we assume the IC
propagation model.

\spara{Datasets, probabilities and methods used.} We use two
real-world networks, whose statistics are reported in
Table~\ref{tab:datasets}.

The first network, called \neth, is the same used in
\cite{ChenWY09, ChenWW10, ChenWW10b}. It is an academic
collaboration network extracted from ``High Energy Physics -
Theory'' section of arXiv\footnote{\url{http://www.arXiv.org}},
with nodes representing authors and edges representing
coauthorship. This is clearly an undirected graph, but we consider
it directed by taking for each edge the arcs in both the
directions. Following \cite{kempe03, ChenWY09, ChenWW10},
we assign probabilities to the arcs in two different
ways: \emph{uniform}, where each arc has probability 0.1 (or probability
0.01) and \emph{weighted
cascade} (WC), i.e, the probability of an arc $(v,u)$ is $p_{v,u} =
1/d_{in}(u)$, where $d_{in}(\cdot)$ indicates in-degree
\citep{kempe03}. Note that WC is a special case of IC where probabilities
on arcs are not necessarily uniform.

\begin{table}[t!]
  \centering
  \begin{tabular}{r|c|c|}
    \multicolumn{1}{c}{} & \multicolumn{1}{c}{\neth} & \multicolumn{1}{c}{\meme}\\ \cline{2-3}
    $\# Nodes$ & 15233 & 7418  \\
    $\# Arcs$ & 62794 & 39170 \\
    $Avg. degree$ & 4.12 & 5.28 \\
     $\# CC$ (strong) & 1781 & 4552 \\
    $max \, CC$ (strong) & 6794 (44.601\%) & 2851 (38.434\%) \\
    clustering coefficient & 0.31372 & 0.06763 \\ \cline{2-3}
  \end{tabular}
  \caption{Networks statistics: number of nodes and directed
  arcs with non-null probability, average degree, number of
  (strongly) connected components, size of the largest one, and clustering coefficient.}\label{tab:datasets}
\end{table}

\begin{table}[h]
\hspace{-10mm}
  \centering
 \begin{small}
\begin{tabular}{r|l|}
 \cline{2-2} \multirow{2}{*}{\textsc{Random}} &  Simply
add nodes at random to the seed set,\\ & until the stopping
condition is met. \\

\cline{2-2} \multirow{2}{*} {\textsc{High Degree}} & Greedily add
the highest degree node to the\\ & seed set, until the stopping
condition is met. \\

\cline{2-2} \multirow{3}{*}{\textsc{Page Rank}} & The popular
index of nodes' importance. \\ & We run it with the same setting used \\
& in \cite{ChenWW10}.\\
%

\cline{2-2}
  \multirow{3}{*}{\textsc{Sp}} &  The shortest-path based heuristic for
  the \\ & greedy algorithm introduced in \\ & \cite{KimuraS06}. \\

\cline{2-2}
 \multirow{3}{*}{\textsc{Pmia}} & The maximum influence arborescence \\
 & method of \cite{ChenWW10} with \\ & parameter $\theta = 1/320$. \\

\cline{2-2} \multirow{3}{*}{\textsc{Greedy}} &  Algorithm
$\MINSEEDSET$ for MINTSS\\ & and Algorithm $\MINSEEDSET^R$
for \\ & MINTIME.
\\ \cline{2-2}
\end{tabular}\end{small}

  \caption{The methods used in our experiments.}\label{tab:methods}
\vspace{-2mm}
\end{table}

The second one, called \meme, is a sample of the social network
underlying the Yahoo! Meme\footnote{\url{http://meme.yahoo.com/}}
microblogging platform. Nodes are users, and directed arcs from a
node $u$ to a node $v$ indicate that $v$ \emph{``follows''} $u$.
For this dataset, we also have the log of posts propagations during
2009. We sampled a connected sub-graph of the social network
containing the users that participated in the most re-posted
items. The availability of posts propagations is significant since it
allows us to directly estimate actual influence.

\begin{figure*}[t!]
\centering
\begin{tabular}{ccc}
\hspace{-8mm}\includegraphics[width=0.35\textwidth]{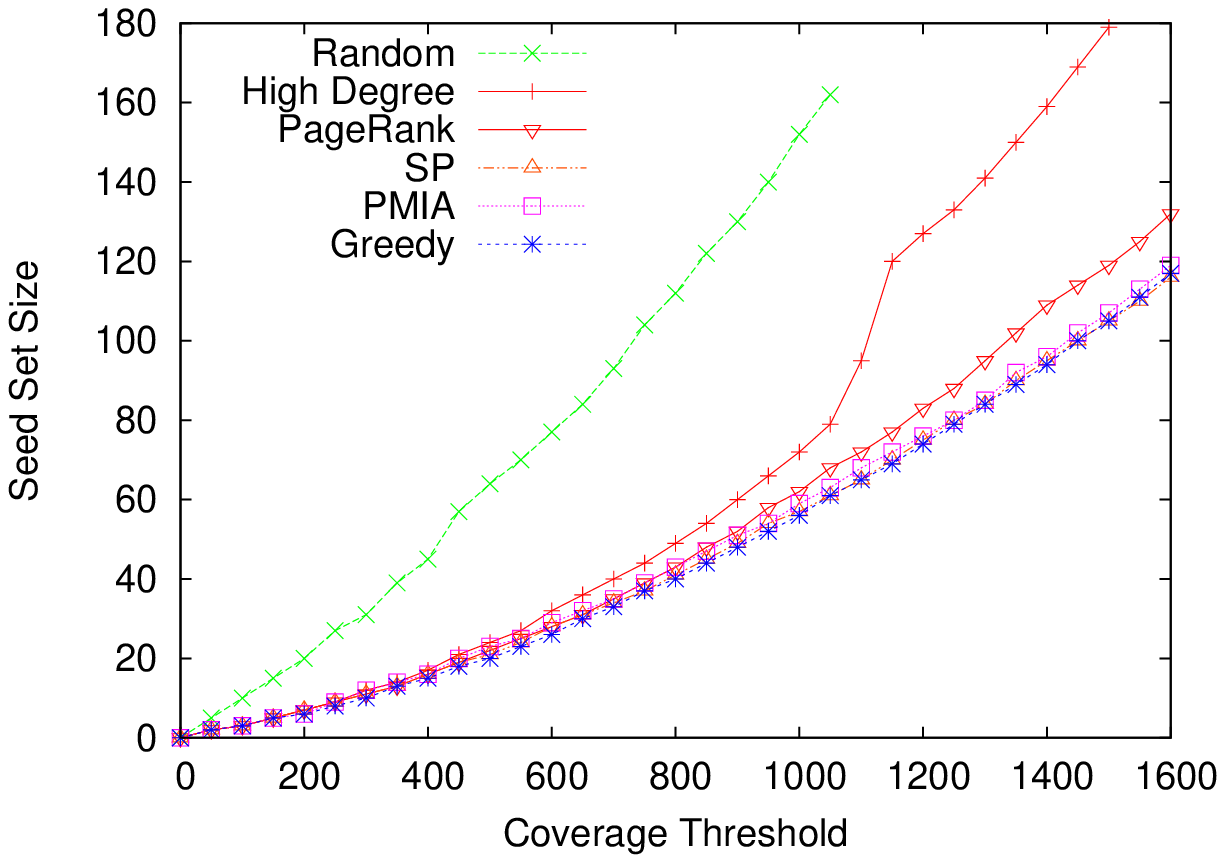}&
\hspace{-4mm}\includegraphics[width=0.35\textwidth]{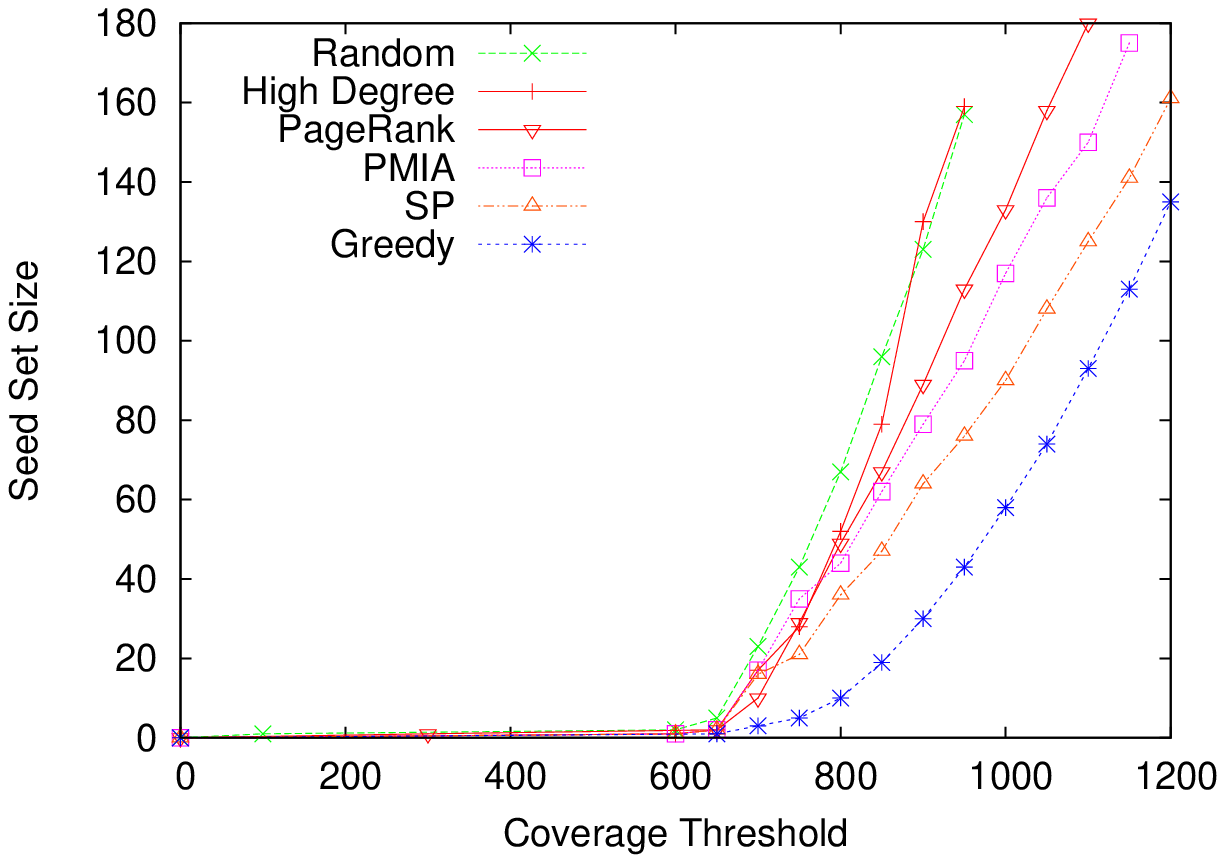}&
\hspace{-4mm}\includegraphics[width=0.35\textwidth]{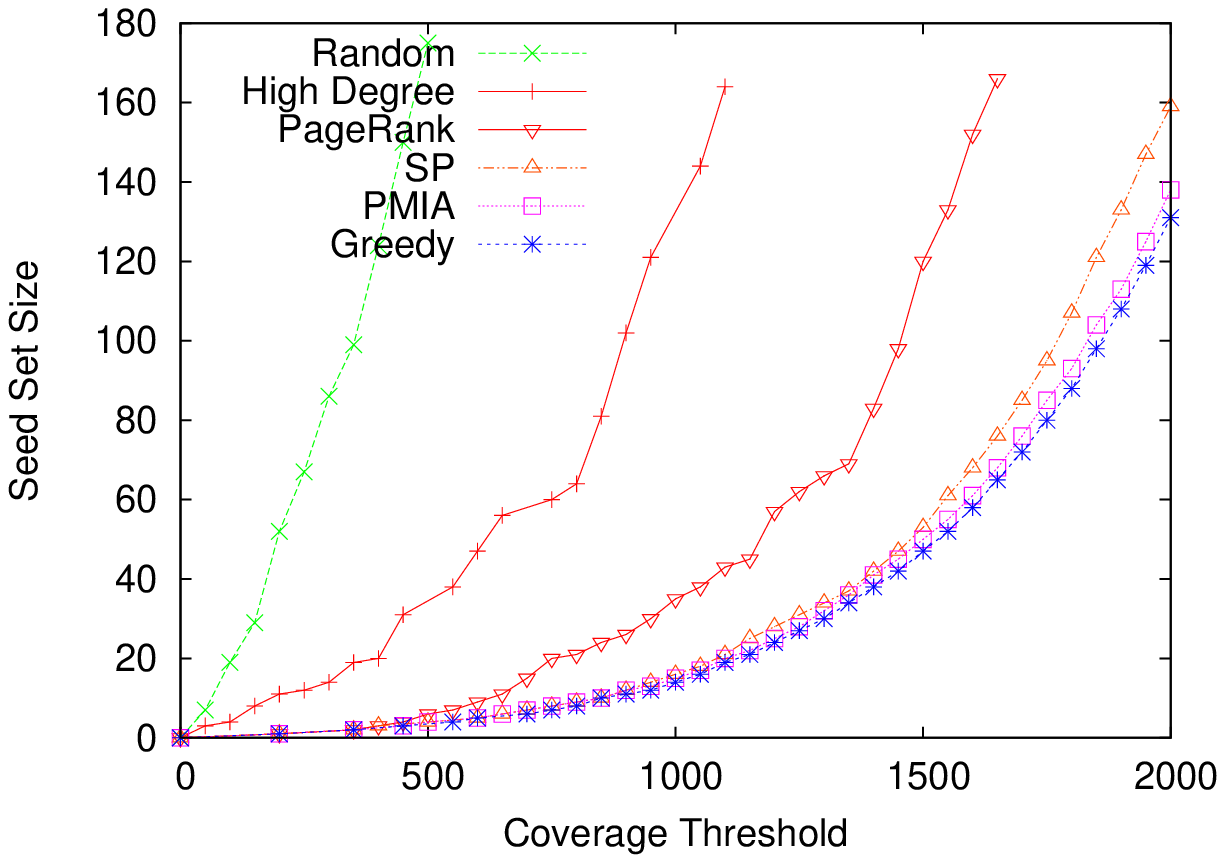}
\\
\hspace{-8mm}(a) \neth - WC  & (b) \neth - uniform & (c) \meme
\end{tabular}
  \caption{Experimental results on MINTSS. }\label{plots:minseed}
\end{figure*}

In particular, here a propagation is defined based on reposts: a
user posts a meme, and if other users like it, they repost it, thus
creating cascades. For each meme $m$ and for each user $u$, we
know exactly from which other user she reposted, that is we have a
relation $repost(u,v,m,t)$ where $t$ is the time at which the
repost occurs, and $v$ is the user from which the information
flowed to user $u$. The maximum likelihood estimator of the probability
of influence corresponding to an
arc is $p_{v,u} =  M_{v2u} / M_{vu}$ where  $M_{vu}$ denotes the
number of memes that $v$ posted before $u$, and $M_{v2u}$ denotes
the number of memes $m$ such that $repost(u,v,m,t)$.

For the sake of comparison, we adapt the state-of-the-art methods
developed for MAXINF (also see
Section \ref{sec:related}) to deal with MINTSS and MINTIME. For most of the techniques the
adaptation is straightforward. The methods that we use in the
experimentation are succinctly summarized in Table
\ref{tab:methods}. It is noteworthy that PMIA is one of the state-of-the-art heuristic
algorithms proposed for MAXINF under the IC model by \cite{ChenWW10}. In all our experiments,
we run 10,000 Monte Carlo simulations for estimating coverage.

\spara{MINTSS -} Our experimental results on the MINTSS problem
are reported in Figure \ref{plots:minseed}. In each of the three
plots, we report, for a given coverage threshold ($x$-axis), the
minimum size of a seed-set (budget, reported on $y$-axis)
achieving such coverage. As \textsc{Greedy} provides the upper
bound on the quality that can be achieved in PTIME, in all the
experiments it outperforms the other methods, with \textsc{Random}
and \textsc{High Degree} consistently performing the worst.

We analyzed the probability distributions of the various
data sets we experimented with. At one extreme is the model
with uniformly low probabilities (0.01).
In \meme, about 80\% of the probabilities are $\leq 0.05$.
In \neth\ WC, on the
other hand,
approximately 83\% of the probabilities are $\geq 0.05$
and about 66\% of the probabilities are $\geq 0.1$. However,
the combination of a power law distribution of node degrees in
\neth\ together with assignment of low probabilities for high degree
nodes (since it's the reciprocal of in-degree) has the effect of rendering
central nodes act as poor influence spreaders. And the arcs
with high influence probability are precisely those that are incident to nodes
with a very low degree. This makes for a low influence
graph overall, i.e., propagation of influence is limited.
Finally, at the other extreme is the model with uniformly high
probabilities (0.1) which corresponds to a high influence graph.

We tested uniformly low probabilities (0.01), and we observed that with
such low probabilities, there is limited
propagation happening: for
instance, in order to achieve a coverage of 150, even the best
method requires more than 100 seeds. This forces the
quality of all algorithms to look similar.

\begin{figure*}[t!]
\centering
\begin{tabular}{ccc}
\hspace{-8mm}\includegraphics[width=0.35\textwidth]{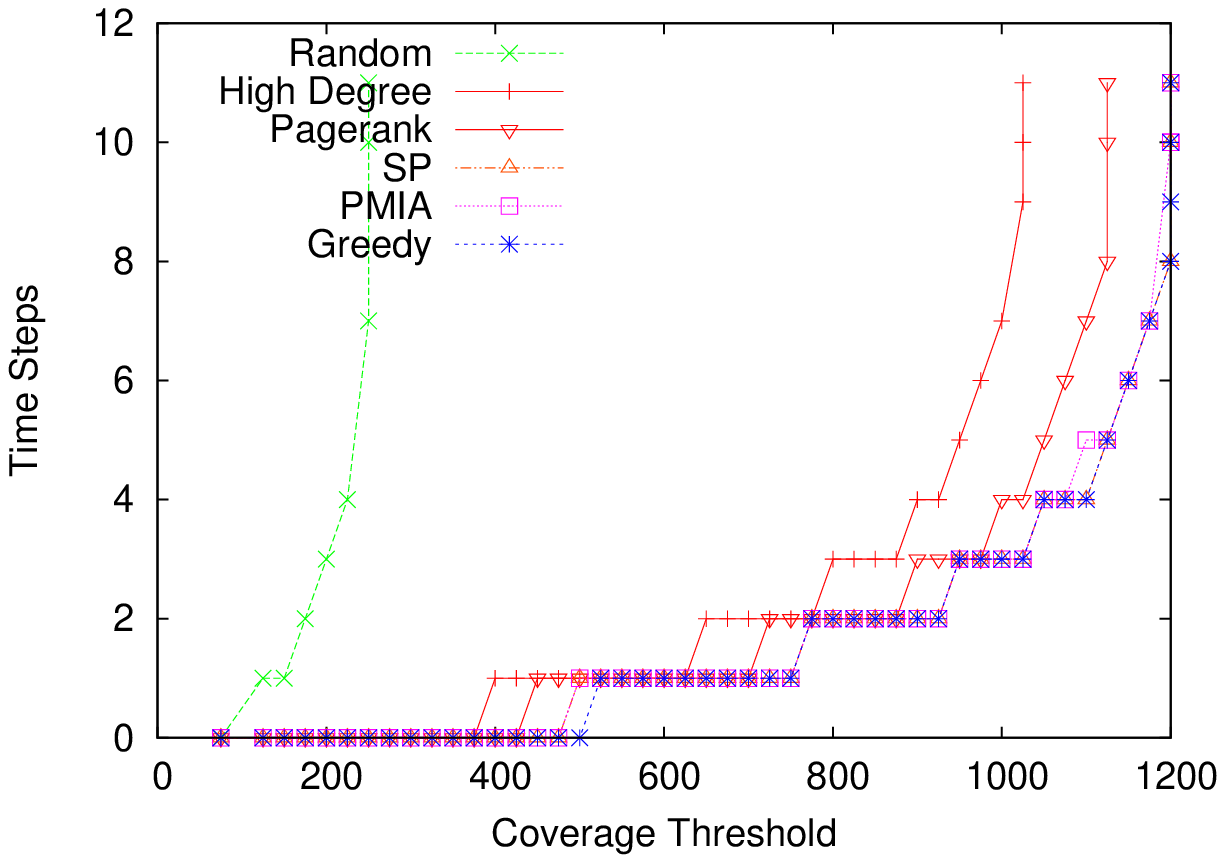}&
\hspace{-4mm}\includegraphics[width=0.35\textwidth]{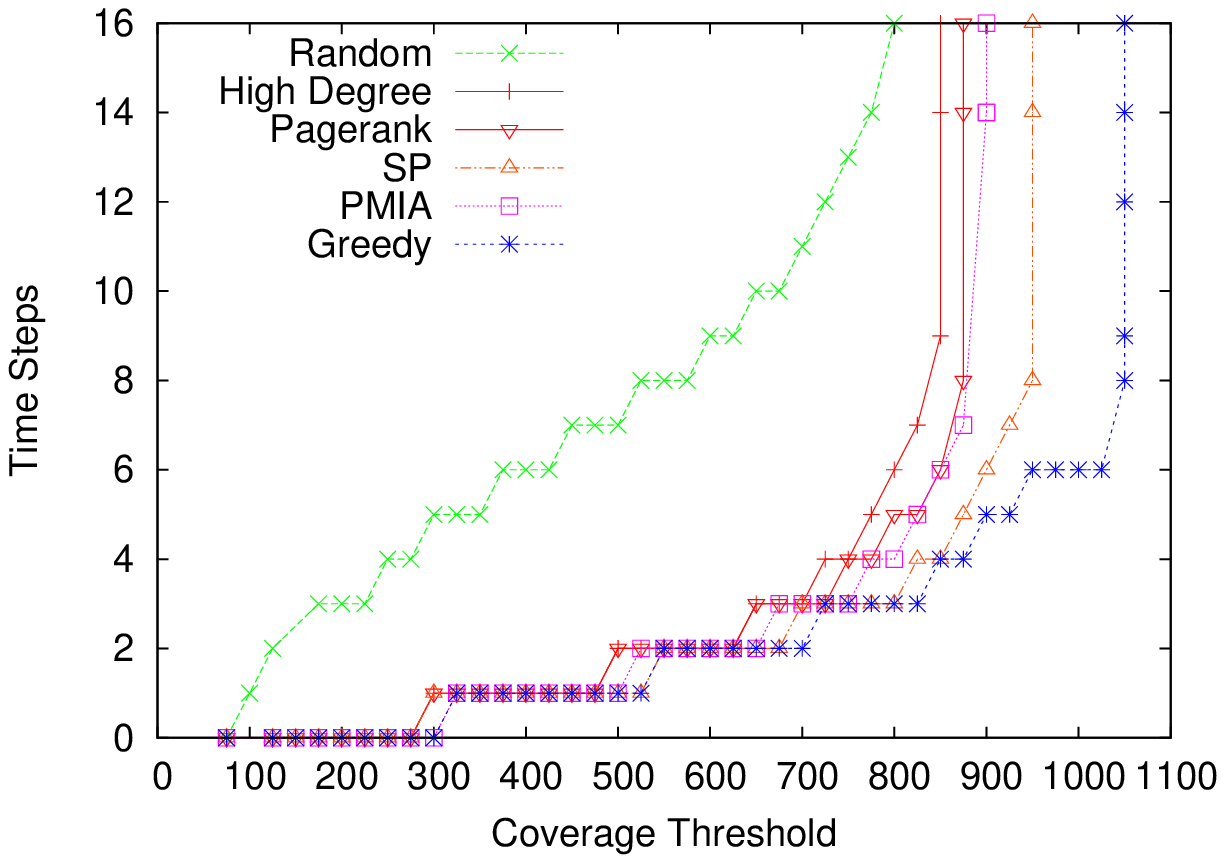}&
\hspace{-4mm}\includegraphics[width=0.35\textwidth]{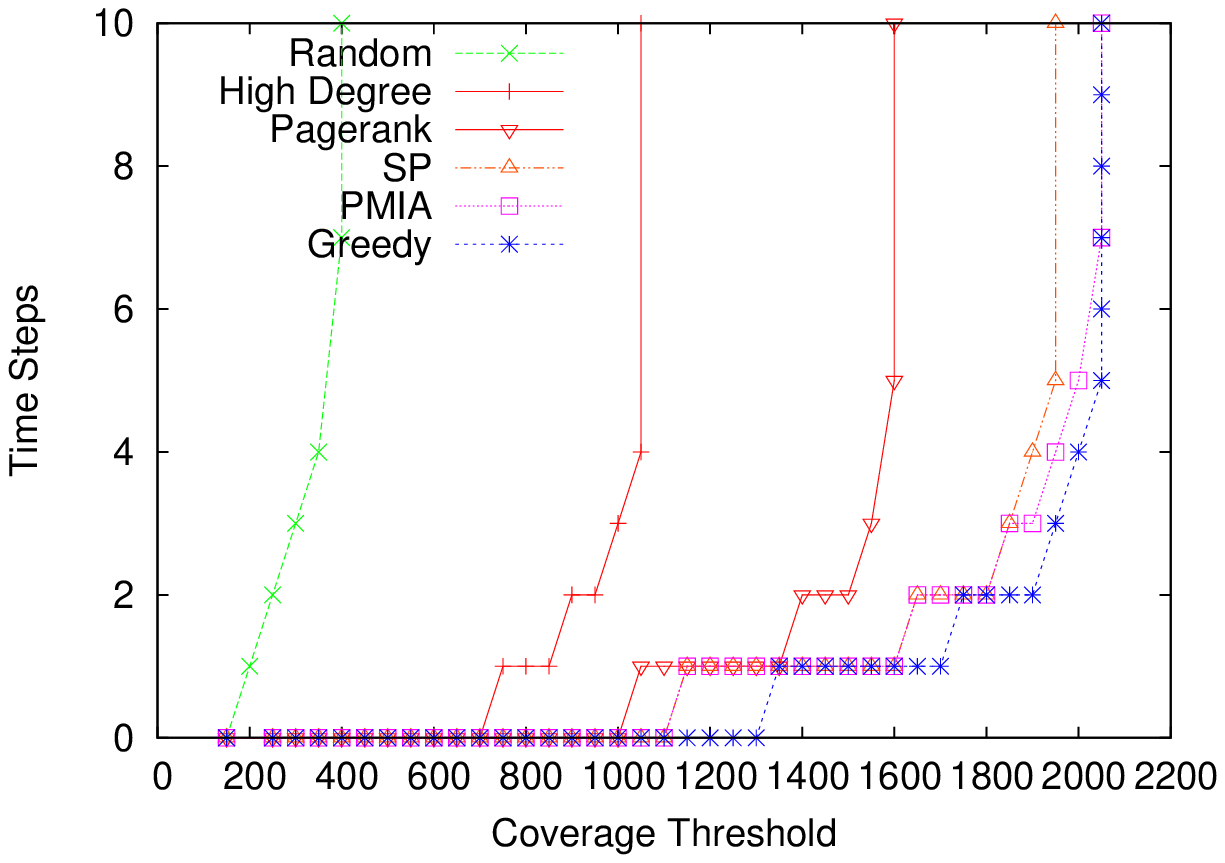}
\\
\hspace{-6mm}(a) \neth - WC, Budget=75 &
(b) \neth - Uniform, Budget=75 & (c) \meme, Budget=150
\end{tabular}
  \caption{Experimental results on MINTIME with fixed budget.}\label{plots:mintime}
\end{figure*}

\begin{figure*}[t!]
\centering
\begin{tabular}{ccc}
\hspace{-8mm}\includegraphics[width=0.35\textwidth]{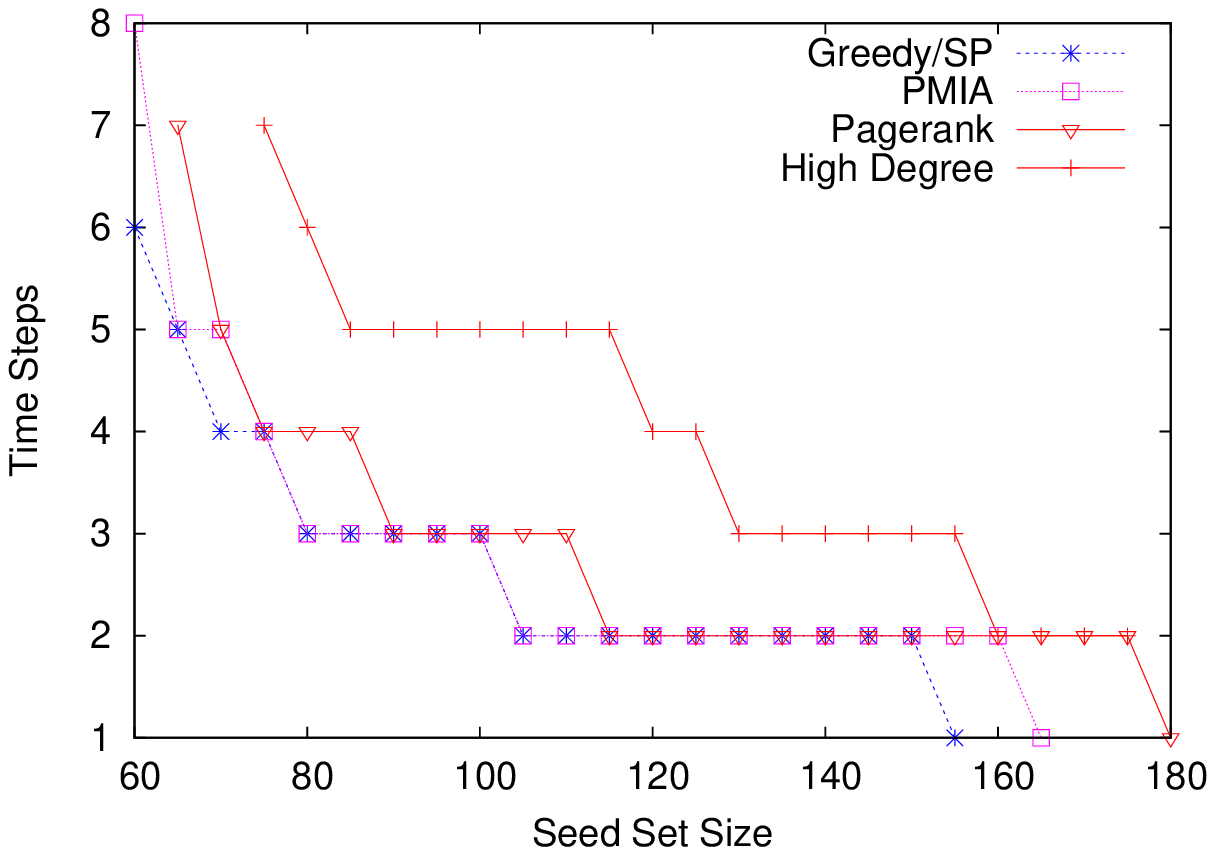}&
\hspace{-4mm}\includegraphics[width=0.35\textwidth]{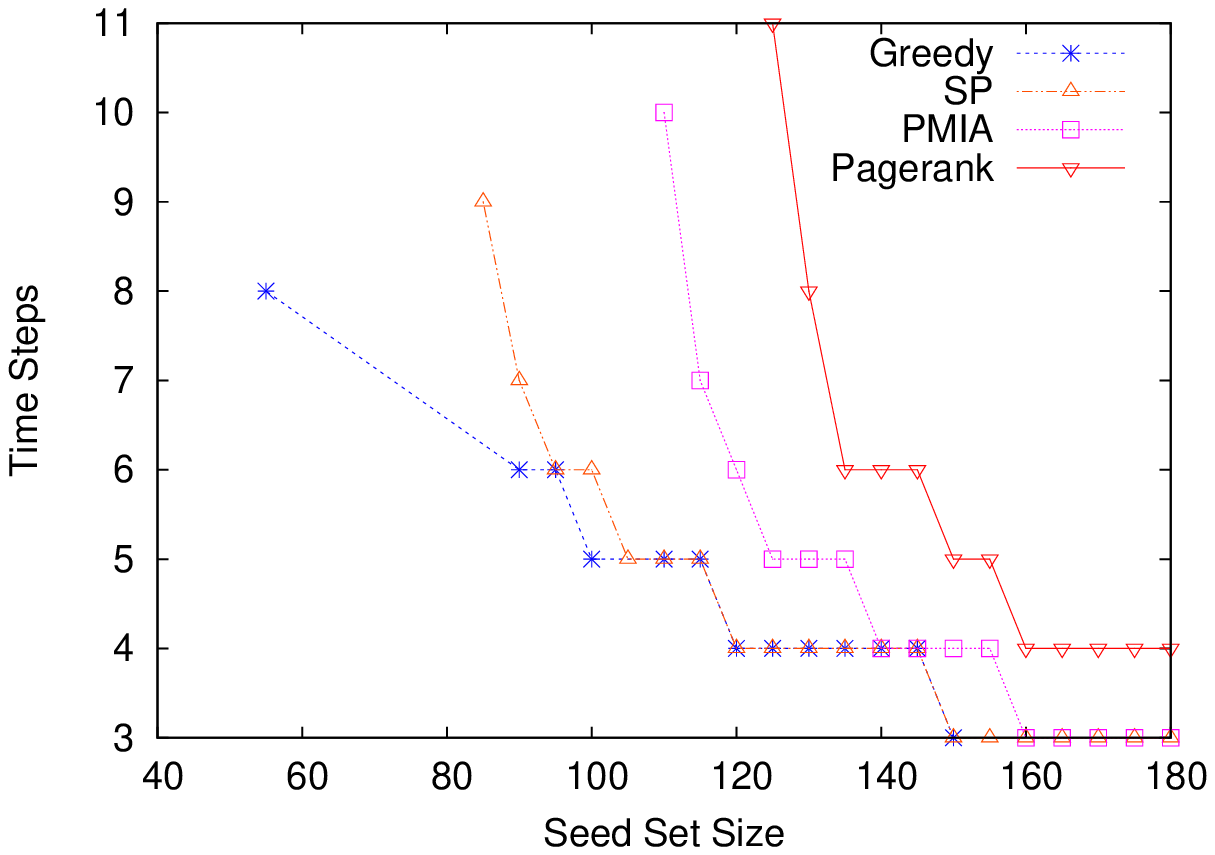}&
\hspace{-4mm}\includegraphics[width=0.35\textwidth]{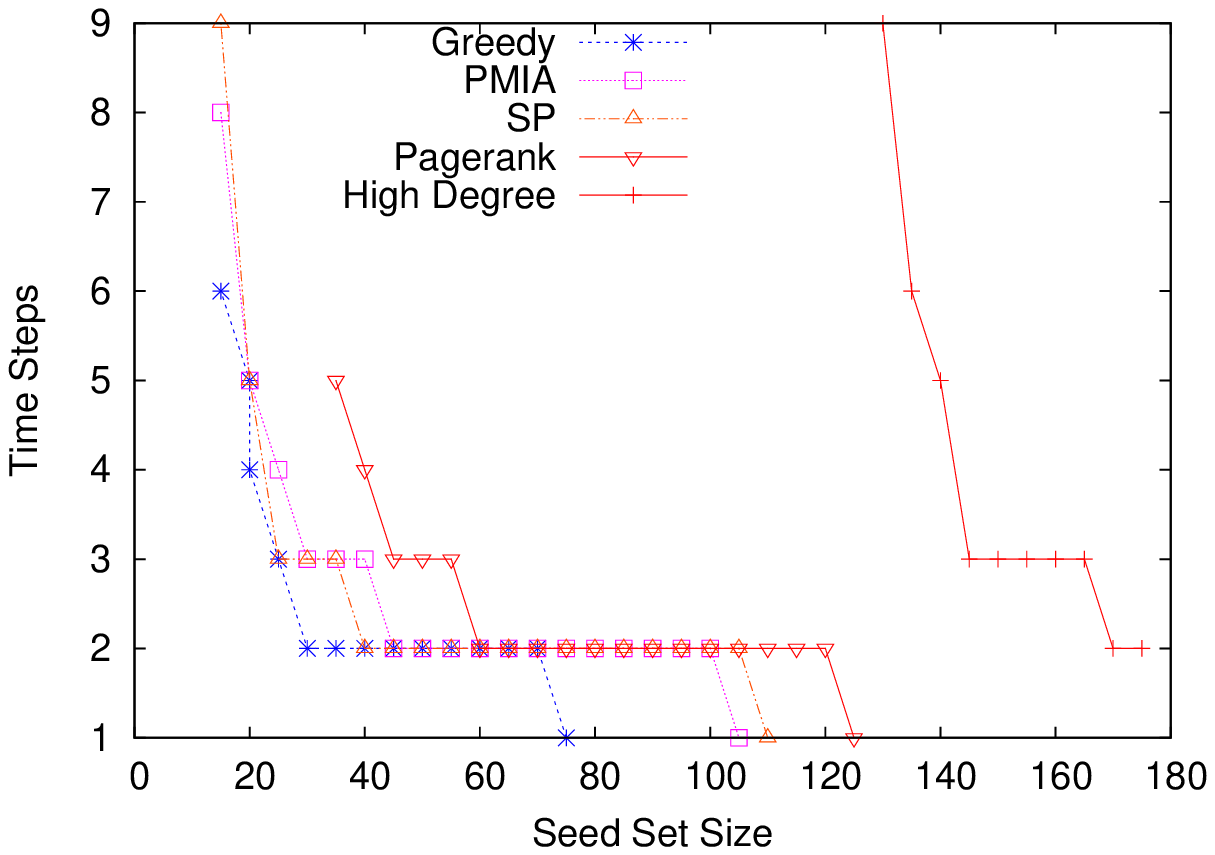}
\\
\hspace{-6mm}(a) \neth - WC, $\eta$=1000 &
(b) \neth - Uniform, $\eta$=1000 & (c) \meme, $\eta$=1000
\end{tabular}
  \caption{Experimental results on MINTIME with fixed Coverage
  Threshold.}\label{plots:mintime2}
\end{figure*}

On data sets where there is a non-uniform mix of low and high probabilities,
but the probabilities being predominantly low, as well as
on data sets corresponding to low influence graphs,
the \textsc{Pmia} method of \cite{ChenWW10} and the \textsc{Sp}
method of \cite{KimuraS06}, originally developed as efficient
heuristics for the MAXINF problem, when
adapted to the MINTSS problem, continue to provide a
good approximation of the results achieved by the \textsc{Greedy}
algorithm (Figure~\ref{plots:minseed}(a), (c)). In these situations,
the Random and HighDegree heuristics provide seed sets much larger than
\textsc{Greedy}. In \neth\ WC (Figure~\ref{plots:minseed}(a)), PageRank
has a performance that is close to the Greedy solution, while in \meme
(Figure~\ref{plots:minseed}(c)), the seed set generated by PageRank is much larger than
Greedy. 
In data sets with uniformly high probabilities (0.1), the gap
between
between \textsc{Greedy} and other heuristics
is substantial
(Figure \ref{plots:minseed}(b)). \textsc{Greedy} can achieve a
target coverage $\eta = 750$, with just $5$ seeds, while
\textsc{Pmia} and \textsc{Sp} need $35$ and $21$ seeds
respectively; similarly \textsc{Greedy} can achieve a target
coverage $\eta = 1000$, with just $58$ seeds, while \textsc{Pmia}
and \textsc{Sp} need $117$ and $90$ seeds respectively.
It is worth noting that Random, HighDegree, and the PageRank heuristic
all generate seed sets much larger than Greedy on this data set.
To sum, the gap between the sizes of the seed sets obtained by the
heuristics one the one hand and the Greedy algorithm on the other,
varies depending on the influence probabilities on the edges. In general,
on graphs with high influence, the gap can be substantial.

\spara{MINTIME -} Our experimental results on the MINTIME problem
are reported in Figures \ref{plots:mintime} and \ref{plots:mintime2}. In
Figure \ref{plots:mintime}, we report, for
a coverage threshold given on the $x$-axis, and a fixed budget (75
for \neth, 150 for \meme), the minimum time steps needed to
achieve such coverage with the given budget ($y$-axis). As
expected, \textsc{Greedy} outperforms all the heuristics. All the
plots show that after a certain time, there is no further gain in
the coverage, indicating the influence decays over time.
Figure~\ref{plots:mintime}(a) compares the various heuristics with
the \textsc{Greedy} on the \neth\ dataset under WC model. On this data
set, \textsc{Pmia}, \textsc{Sp} and \textsc{Greedy} exhibit
comparable performance. The \textsc{Pagerank} heuristic comes
close to them. 

Figure~\ref{plots:mintime}(b) shows the results for the \neth\ dataset
under IC model with uniform probability 0.1. Here, \textsc{Greedy}
outperforms all the other heuristics. For instance, when coverage
threshold $\eta$ is 900 and budget is 75, \textsc{Greedy} achieves
the coverage in 5 time steps, and \textsc{Sp} in 6 time steps,
\textsc{Pmia} in 14 time steps. \textsc{Random}, \textsc{High
Degree} and \textsc{Pagerank} fail to find a solution.
Similarly, when coverage threshold is 1000 and budget is 75,
\textsc{Greedy} achieves the coverage in 6 steps whereas all
other heuristics fail to find a solution with this
coverage.

Finally, Figure~\ref{plots:mintime}(c) shows the results on \meme\
dataset. As we increase the target coverage, the other heuristics
fail to give a solution, one by one. Beyond $\eta=1600$, all but \textsc{Sp},
and \textsc{Pmia} fail and beyond $\eta=2000$, all but
\textsc{Pmia} fail. On this data set,
\textsc{Pmia}  provides
a good approximation  to the performance of \textsc{Greedy}.

In Figure \ref{plots:mintime2}, we fix the coverage threshold
($\eta=1000$ for all the plots). The plots show the minimum time steps
needed to achieve the coverage w.r.t.~different seed set sizes (budget).
In all the cases, \textsc{Random} fails to find a solution and hence is
not shown in the plots. The performance of the \textsc{High
Degree} algorithm is poor as well and it fails to find a solution in
case of $\neth$ with uniform probabilities 0.1.
As expected, \textsc{Greedy} outperforms all the heuristics and provides
us the lower bound on time needed to achieve the required coverage with
a given budget.

Overall, we notice that the performance quality of all other heuristics
compared to \textsc{Greedy} follows a similar pattern to that
observed in case of MINTSS: as the graph changes from a low influence
graph to a high influence graph, the heuristics' performance drops substantially
compared to \textsc{Greedy}.

Another key takeaway from the MINTIME plots is the following. {\sl For a
given budget, as observed above, the choice of the seed set plays a key
role in determining whether a given coverage threshold can be reached or not,
no matter how much time we allow for the influence to propagate.
Even if the given coverage threshold is achieved, the choice of the
seed set can make a big difference to
the number of time steps in which the coverage threshold is reached.}
Often, for a given budget, relaxing the coverage threshold can
dramatically change the propagation time. E.g., In
Figure~\ref{plots:mintime}(a) (budget fixed to 75), while \textsc{Greedy}
takes 8 time steps to achieve a coverage of 1200, when we relax the threshold to 1100,
the propagation time decreases by $50$\%, that is, to just 4 time steps.
A similar phenomenon is observed when the budget is boosted w.r.t. a fixed
coverage threshold.
For instance, in Figure~\ref{plots:mintime2}(c), while using 15 seeds, \textsc{Greedy} takes 6 time
steps to achieve a coverage of 1000, it achieves the same coverage by
$30$ seeds in $33$\% of the time, that is, in 2 time steps.
These findings further highlight the importance of the MINTIME
problem.

%% file: discussion.tex
In this paper, we study two optimization problems in social
influence propagation: MINTSS and MINTIME.  We present a
bicriteria approximation for MINTSS which delivers a seed set larger than the optimal
seed set by a logarithmic factor $(1+\ln(\eta/\epsilon))$, that achieves a
coverage of $\eta - \epsilon$, which falls short of the coverage threshold
by $\epsilon$. We also show a generic tightness result that
 indicates improving the above approximation factor is likely to be hard.

Turning to MINTIME, we give a greedy algorithm that provides a tricriteria
approximation when allowed a budget overrun by a factor of
$(1+\ln(\eta/\epsilon))$ and a coverage shortfall by $\epsilon$,
and achieves the optimal propagation time under these conditions. We also provide hardness
results for this problem.
We conduct experiments on two real-world networks to compare the
quality of various popular heuristics proposed in a different
context (with necessary adaptations) with that the greedy approximation algorithms. Our results
show that the greedy algorithms outperform the other
methods in all the settings (as expected)  but depending on the characteristics
of the data, some of the heuristics perform competitively. These
include the recently proposed heuristics \textsc{Pmia} \cite{ChenWW10} and
\textsc{Sp} \cite{KimuraS06} which we adapted to MINTSS and MINTIME.

Several questions remain open, including proving optimal
approximation bounds for MINTSS and MINTIME, as well as complexity
results for these problems under other propagation models.


